\documentclass[fleqn,usenatbib]{rasti}

\usepackage{newtxtext,newtxmath}
\usepackage[T1]{fontenc}

\DeclareRobustCommand{\VAN}[3]{#2}
\let\VANthebibliography\thebibliography
\def\thebibliography{\DeclareRobustCommand{\VAN}[3]{##3}\VANthebibliography}

\usepackage{graphicx}	% Including figure files
\usepackage{amsmath}	% Advanced maths commands
\usepackage{orcidlink}
\usepackage{anyfontsize}
\usepackage{soul}
\usepackage{amsmath}	% Advanced maths commands
\usepackage{siunitx} % For \unit and other SI unit formatting
\newcommand{\eps}{{e$^-$\,pix$^{-1}$\,s$^{-1}$}}
\newcommand{\tess}{\textit{TESS}}
\newcommand{\hst}{\textit{HST}}
\newcommand{\gaia}{\textit{Gaia}}
\newcommand{\kepler}{\textit{Kepler}}
\newcommand{\plato}{\textit{PLATO}}

\usepackage[normalem]{ulem}

\title[Marana CMOS Camera]{High-Precision Photometry with a scientific CMOS Camera: I Lab Testing of the Marana camera}

\author[Ioannis Apergis et al.]{
Ioannis~Apergis$^{\orcidlink{0009-0004-7473-4573}}$,$^{1,2}$\thanks{E-mail: Ioannis.Apergis@warwick.ac.uk (IA)}
Daniel~Bayliss$^{\orcidlink{0000-0001-6023-1335}}$,$^{1,2}$
Leonidas~Asimakoulas$^{\orcidlink{0000-0003-3676-5606}}$,$^{4}$
Paul~Chote,$^{2,3}$
James~McCormac$^{\orcidlink{0000-0003-1631-4170}}$,$^{1,2,3}$
\newauthor
Morgan~A.~Mitchell$^{\orcidlink{0009-0004-6130-7775}}$,$^{1,2}$
Sam~Gill$^{\orcidlink{0000-0002-4259-0155}}$,$^{1,2}$
Philip~G.~Steen$^{4}$
and Peter~J.~Wheatley$^{\orcidlink{0000-0003-1452-2240}}$$^{1,2}$
\\
$^{1}$Centre for Exoplanets and Habitability, University of Warwick, Gibbet Hill Road, Coventry CV4 7AL, UK\\
$^{2}$Department of Physics, University of Warwick, Gibbet Hill Road, Coventry CV4 7AL, UK\\
$^{3}$Centre for Space Domain Awareness, University of Warwick, Gibbet Hill Road, Coventry CV4 7AL, UK\\
$^{4}$Andor Technology Ltd, Springvale Business Park, 7 Millennium Way, Belfast BT12 7AL, UK
}

\date{Accepted XXX. Received YYY; in original form ZZZ}

\pubyear{2025}

\begin{document}
\label{firstpage}
\pagerange{\pageref{firstpage}--\pageref{lastpage}}
\maketitle

% Abstract of the paper
\begin{abstract}
Scientific CMOS cameras are becoming increasingly prevalent in modern observational astronomy. We assess the ability of CMOS image sensors technology to perform high-precision photometry with a detailed laboratory characterization of the Marana 4.2BV-11 CMOS camera. We characterise the camera in the Fastest Frame Rate (FFR) and High Dynamic Range (HDR) modes. Our evaluation includes read noise, dark current, photo response and dark signal non-uniformities, quantum efficiency and window transmittance. The read noise is found to be 1.577\,e$^-$ for the FFR mode. For the HDR mode the read noise floor is measured to be 1.571\,e$^-$ for signal levels below approximately 1800\,e$^-$. The bias level shows dark signal non-uniformities with values of 0.318\,e$^-$ and 0.232\,e$^-$ for FFR and HDR mode, respectively. Pixel well capacity reached 2366 e$^-$pix$^{-1}$ for the FFR mode and 69026 e$^-$pix$^{-1}$ with a dynamic range of 93\,dB for the HDR mode. The camera demonstrates good linearity, yielding linearity errors of 0.099\,\% for FFR mode and 0.122\,\% for HDR mode. The uniformity across the image arrays show a photo response non-uniformity of 0.294\,\% for the FFR mode and 0.131\,\% for the HDR mode. The dark current displays a noticeable glow pattern, resulting in mean dark current levels of $1.674\pm0.011$\, \eps\, for the FFR mode and $1.617\pm0.008$\,\eps\, for the HDR mode at a constant temperature of -25\,$^\circ$C. We measured the quantum efficiency across the visible spectrum, with a peak of of >95\,\% at 560\,nm. Our tests indicate that the Marana CMOS camera is potentially capable of performing precise photometry. 
\end{abstract}
% Include between one and six keywords.

%%%%%%%%%%%%%%%%%%%%%%%%%%%%%%%%%%%%%%%%%%%%%%%%%%

%%%%%%%%%%%%%%%%% BODY OF PAPER %%%%%%%%%%%%%%%%%%

\section{Introduction}
\label{introduction}
Charge-Coupled Devices \citep[CCDs,][]{boyle1970charge} are almost universally used for optical astronomical imaging. Examples relating to high precision photometry include the Hubble Space Telescope \citep[\hst]{1998SPIE.3356..234F},
\kepler\ \citep{2010Sci...327..977B}, Transiting Exoplanet Survey Satellite \citep[\tess]{2015JATIS...1a4003R}, \gaia\ \citep{2016A&A...595A...1G}, and PLAnetary Transits and Oscillations \citep[\plato]{rauer2016plato}. CCD cameras feature outstanding performance due to their high sensitivity, low read noise, stability, and good pixel-to-pixel uniformities. Additionally, developments in the sensor architecture such as deep depletion sensors \citep{1480428, peckerar1981x, 1985ITED...32.1525T, gimeno2016sky}, and back-side illuminated structures, provide exceptional quantum efficiency across a wide range of wavelengths. With thermoelectric or cryogenic cooling, the dark current of CCDs can be reduced to just a few electrons per pixel per hour. 

Another type of image sensor developed alongside CCDs over the years is Complementary Metal-Oxide Semiconductor (CMOS). Unlike CCDs with a passive pixel sensor array, CMOS are active pixel sensors, where each pixel contains Field-Effect Transistors (MOSFETs) to convert the charge to voltage on a pixel-by-pixel basis. The individual in-pixel circuitry traditionally resulted to higher noise compared to a CCD with a single amplifier architecture \citep{1999SPIE.3649..177T}. Nonetheless, the rapid readout speeds and low power consumption make CMOS sensors ideal for consumer markets. In contrast to CCDs, which are now primarily limited to scientific applications, the significant demand within the commercial market and applications utilizing camera technology has driven manufacturers to develop increasingly advanced CMOS devices by minimizing pixel architectures down to sub-micron level \citep{vu2008design, 9365751, uchiyama202140}. The high production volume of CMOS detectors has made them significantly cheaper and more readily available than CCD detectors. Recent laboratory and on-sky testing results show that CMOS cameras have the potential to compete with CCDs cameras in certain astronomical applications \citep{jorden2014e2v, 2019AN....340..638K, 2023sndd.confE...3A, 2025arXiv250312449X, 2025arXiv250200101L}.

Today, scientific CMOS \citep[hereafter CMOS;][]{coates2009scientific} cameras have found significant usage in astronomy \citep{qiu2013evaluation, 2019AN....340..638K, wang2020test, ardilanov2021high, 2021RAA....21..268Q, stefanov2022cmos, townsend2023electro, 2023sndd.confE...3A, 2024SPIE13103E..0RK} due to their fast readout speed, allowing them to record in short exposures and to maximise the duty cycle. CMOS cameras are also available in large sensor sizes, resulting in a wide field of view that is mandatory for sky surveys, allowing for simultaneous observations of many stars in a single image \citep{2020SPIE11525E..2IW, 2025arXiv250200101L}. In contrast to the Analogue-to-Digital Converter (ADC) of CCD cameras, the parallel readout architecture of CMOS cameras, an ADC per column, digitises the voltage signal faster while preserving low power consumption. Thanks to the pinned photodiode pixel architecture and the on-chip amplifier per pixel, CMOS image sensors can achieve low read noise while conserving high frame rates and extended dynamic range \citep{inoue1999low, 839709, fossum2014review}. The large number of amplifiers and parallel readout path in CMOS sensors allow each pixel to be processed at lower frequencies, resulting in low read noise. In contrast, CCDs require relatively higher sampling frequencies to process a large number of pixels and therefore lead to higher read noise \citep{628823, 1485947}.

Unlike front-illuminated CMOS sensors, recent developments such as back-illuminated sensors \citep{wuu2009manufacturable, wang20154m}, have pushed quantum efficiency values above 90\% in visible wavelengths \citep[e.g.][]{2019AN....340..638K, 2023sndd.confE...3A}.
CMOS cameras have been proposed for implementation in various astronomical projects such as ATLAS \citep{2018PASP..130f4505T}; ASTEP \citep{2020SPIE11447E..0OC}, Argus Optical Array \citep{2022PASP..134c5003L} and Mini-MegaTORTORA \citep{2022aems.conf..156K}. Under the advanced deployment of optical blocking filters and films, such cameras have also been implemented for X-ray space missions for example in Einstein Probe \citep[e.g the Wide Field X-ray Telescope (WXT) in][]{2018SPIE10699E..25Y} and have also been commissioned for the Earth 2.0 \citep{2022arXiv220606693G} and Theseus \citep{2020SPIE11444E..8YM} future missions. 

One of the key astronomical applications of large-format digital detectors is time-series photometry. In particular, large-format CCD detectors have been used to simultaneously monitor the brightness of thousands of stars over wide areas of the sky. Surveys such as WASP \citep{2006PASP..118.1407P}, Kepler \citep{2010Sci...327..977B}, \tess\ \citep{2015JATIS...1a4003R} and NGTS \citep{2018MNRAS.475.4476W} all use large format CCD detectors to achieve high precision time-series photometry in order to discover transiting exoplanets.

In this study, we aim to investigate whether modern CMOS detectors can achieve high precision photometry to a level competitive with CCD detectors. To do this, we characterise and test a modern CMOS camera, the Marana 4.2BV-11 (hereafter Marana) developed by Oxford Instruments Andor\footnote{\url{https://andor.oxinst.com/}}, under controlled laboratory conditions. In future work, we will extend this characterisation to on-sky observations at the the NGTS facility \citep{2018MNRAS.475.4476W} at ESO's Paranal Observatory in Chile. The Marana has a low noise floor, high sensitivity, linear response and extended dynamic range. These properties make the Marana a good candidate for high-precision time series photometry. The Marana camera offers several advantages over other CMOS cameras, such as larger dynamic range and pixel size compared to the QHYCCD cameras used in \citet{2023sndd.confE...3A}, which is beneficial for photometry of bright stars. Additionally, the Marana delivers lower read noise than typical CCDs while maintaining significantly faster readout speeds at full frame rates.
 
Three significant studies have examined the performance of the Marana camera. In \cite{2019AN....340..638K}, key performance characteristics were evaluated, including photon transfer curves, linearity, dark current, read noise, spatial autocorrelation, random telegraph signal analysis, and on-sky tests. The study by \cite{2021RAA....21..268Q} compared the Marana with the iKon-L camera across various readout modes, focusing on photon transfer analysis, linearity, read noise, dark current, and on-sky observations. Most recently, \cite{2024SPIE13103E..0RK} investigated the Sona-11 (a rebranded version of the Marana with the same sensor) along with several other CMOS cameras.

In this work, we evaluate the performance of the Marana camera using a series of established and extended methodologies in a similar manner to \citet{2019AN....340..638K, 2021RAA....21..268Q, 2024SPIE13103E..0RK}. Specifically, we measure the photon transfer curve, linearity, read noise, and dark current. In addition to the existing literature, we assess the dark signal and photo response non-uniformities (DSNU and PRNU), and analyse the temporal noise characteristics of columns and rows under dark conditions. The dark current and glow are characterized across a range of operating temperatures. We also present quantum efficiency measurements across a range of temperature of the Marana camera. Finally, to distinguish the sensor quantum efficiency over that of the camera,  we report transmission data of the camera optical front window. Additionally, a detailed description of the pixel readout process of the Marana cameras is included.

In this first paper, we perform a thorough characterisation following the latest EMVA-1288 version 4.0 standards \citep{jahne2010emva}. In Section~\ref{marana} we describe the Marana camera and its CMOS sensor. The laboratory set-up is provided in Section~\ref{sec:equipment}. We explain the techniques and the methods used in Section~\ref{methods}. In Section~\ref{results} we present our results from laboratory testing and the discussion of our findings. Lastly, Section~\ref{conclusion} outlines our conclusions.

\section{The Marana Camera}
\label{marana}

\subsection{The Marana 4.2BV-11 Camera}
\label{instrumentation}
The Marana camera\footnote{\url{https://andor.oxinst.com/products/CMOS-camera-series/marana-CMOS}} is equipped with a thinned backside-illuminated sensor from GPixel\footnote{\url{https://www.gpixel.com/en/index.html}}, specifically the GSENSE400BSI model \citep{lens.org/144-331-726-961-627}. The sensor features a dual-amplifier structure with 4.2\,megapixels each 11\,$\unit{\um}$ in size resulting in a 32\,mm diagonal. The backside illuminated thinned depletion region results in exceptional quantum efficiency in the visible wavelength range. We have selected the Visible-Near Infrared (VIS-NIR) Enhanced Unwedged window, which provides high transmission at the visible and red wavelengths (see Section~\ref{sec:QE_method}). The sensor is enclosed within a vacuum chamber that allows for deep cooling and protects the sensor from condensates. The thermoelectric cooling system ensures effective sensor temperature control, maintaining temperatures down to -25$^\circ$C for ambient temperatures up to 30$^\circ$C. Additionally, water cooling allows for the sensor to comfortably cool down to -45$^\circ$C for ambient temperatures up to 30$^\circ$C. The Marana weighs 3\,kg (6.61\,lbs), and an F-mount with a flange focal distance of 46.5\,mm and a choke diameter of 43.3\,mm is positioned in front of the sensor to maximize the usable field of view of the sensor. This configurations supports beams with f-number of approximately 1.07. Faster beams may cause vignetting. The camera consumes less than 50\,W power and comes with USB3 interface, capable of connectivity up to 5 Gbps. For a summary of the specifications of the Marana camera, see \autoref{tab:specs}.

\begin{table}
\centering
\caption{Specifications of the Andor Marana CMOS camera, as provided on the camera's performance sheet.}
\label{tab:specs}
\begin{tabular}{lllcc}
\hline 
\hline
\textbf{Camera}                &  &  &     \textbf{Andor CMOS}           \\ \hline  
Model                          &  &  &     Marana-4BV11                   \\
Chip                           &  &  &     GSENSE400BSI                   \\
Active pixels                  &  &  &     2048 $\times$ 2048                    \\
Pixel size                     &  &  &     11 $\times$ 11\,$\unit{\um}$                  \\
Sensor size                    &  &  &     22.5 $\times$ 22.5\,mm                 \\
Peak QE                        &  &  &     95\%  @570\,nm                  \\
Readout modes                  &  &  &     100MHz @16-bit, 200MHz @12-bit \\
Shutter type                   &  &  &     Rolling shutter                \\
Gain, e$^{-}$/ADU              &  &  &     1.12 @16-bit, 0.63 @12-bit     \\
Pixel Well Depth, e$^{-}$      &  &  &     67995 @16-bit, 2679 @12-bit    \\
Linearity                      &  &  &     \textgreater 99.7\,\%           \\
Read Noise, e$^{-}$            &  &  &     1.57 @16-bit, 1.58 @12-bit     \\
Dark current, e$^{-}$/pix/s    &  &  &     0.41 @ -45\,$^{\circ}$C          \\ \hline           
\end{tabular}
\end{table}

\subsection{Low Gain and High Gain channels}
\label{sec:hglg}
The CMOS pixel structure of the Marana camera is based on field effect transistors with functionalities such as amplification and charge transfer. For each integration the charge will be transferred in a sense node of a certain capacitance before it is amplified and readout. Additionally, the sensor is based on a 4 Transistor (4-T) pinned photodiode pixel structure \citep{ma20154mp, lens.org/144-331-726-961-627, wang20154m}. On top of the standard 4-T structure this sensor comprises an additional transistor which is switched off/on in order to enhance the sensitivity for low/high illumination levels, respectively, resulting in a High Gain (HG) and Low Gain (LG) capacitance. When the HG is assigned, the sense node will have a small capacitance leading to high gain and therefore low read noise. On the other hand, the sense node will have a large capacitance when the LG is employed. 

During the readout sequence, each HG and LG pixel-level pre-amplifiers are read out through HG or LG column amplifiers positioned at the top or bottom of the sensor \citep{ma20174}. For low light conditions the signal is amplified using the HG pixel-level setting, whereas in brighter light conditions the LG pixel-level setting is used. The HG pixel-level setting is utilized to achieve low read noise \citep{kawai2004noise}, although it comes with a limited full well capacity. The LG pixel-level setting is utilized to achieve an extended full well capacity, but comes at the cost of higher read noise \citep{ma20154mp}. To construct an image, the HG and LG channels can be used in two different imaging modes which we describe in the following section.

\subsection{High Dynamic Range and Fastest Frame Rate imaging modes}
\label{sec:modes}
The Marana camera offers two distinct imaging modes: the High Dynamic Range (HDR) and the Fastest Frame Rate (FFR). Both modes have been characterised in this study to determine the most suitable one for our imaging requirements.

The HDR mode employs the HG and LG pixel-level pre-amplifiers, to capture an extended dynamic range for the image. Each pixel produces two signals (HG and LG), which are passed to the HG and LG column amplifiers, respectively, and digitised by a dual 12-bit column ADC \citep{ma2013low, tang2020high, lou2023over}. A selection algorithm is implemented within the camera's Field-Programmable Gate Array (FPGA) to identify the most appropriate channel output for each recorded pixel value, based on the saturation level of the HG channel. The LG pixel values are then scaled to match the gain of the HG channel and the frame is being reconstructed based on the previous selection algorithm resulting in a 16-bit HDR image \citep{ma20154mp}. The HDR mode operates at 100\,MHz pixel readout rate, resulting in 24 full frames-per-second (fps).

For the FFR mode, the LG column amplifier is being repurposed to operate at the same gain setting as the HG, resulting in two HG column amplifiers. Odd numbered rows are being read using the first HG column amplifier, and even numbered rows are being read using the second HG column amplifier. This results in the readout time being twice as fast as the HDR mode, and therefore the sensor outputs pixels at a frequency of 200\,MHz and 48 full fps. The FFR configuration is also being digitised with 12-bit ADCs, therefore the dynamic range is limited. The specifications of HDR and FFR modes for the Marana camera are shown in \autoref{tab:specs}, which is based on the technical performance details shipped with the camera.

\subsection{Amplifier Glow}
\label{sec:glow}
CMOS cameras exhibit amplifier glow, which may be caused by photons that are emitted by the electronics situated in the periphery of the sensor or from the excess of infrared photons emitted from the Source Follower amplifier \citep{1985SPIE..570....7J}. Additionally, based on \cite{wang20154m} glow structures can be associated with other electronic circuits. Amplifier glow can be a major issue for certain CMOS devices and can emerge as the primary noise source in low-light imaging, thereby compromising their sensitivity. The impact is particularly notable in various astronomical applications utilizing CMOS sensors, spanning optical \citep{wang2020test}, x-ray \citep{townsend2023electro}, and ultraviolet \citep{greffe2022characterization} astronomy. Several approaches have been suggested to address amplifier glow, with one notable method involving the deactivation of electronic amplifiers not in use during exposure \citep{greffe2022characterization}. 

The Marana camera comprises an on-board algorithm to subtract a pre-computed map of the glow from the image. This map is constructed by acquiring a long exposure dark image, in order to capture the glow structure. The dark signal is subtracted from this image leaving behind only the signal from the glow contribution. This reference map is then scaled according to the user requested exposure time and subtracted on a pixel-per-pixel basis. This is known as an Anti-Glow correction. In our testing, we use the Marana camera with and without the Anti-Glow correction. Additionally, strategies to remove the glow pattern have been proposed, including the use of spatially smooth matrices to subtract the effect of amplifier glow during post-acquisition processing \citep{zhang2023glow}. In this work, we consider the dark signal as a mixture of the true dark signal and glow because both obey shot noise properties and cannot be distinguished from each other.

\subsection{Spurious Noise Filter}
\label{sec:spurious}
CMOS detectors can show a number of features that lead to spurious read-out levels at the pixel level. One example is Random Telegraph Signal \citep[RTS,][]{4154408} which can randomly affect any pixel of the sensor. When pixel integration is completed, the charge will be transferred from the photodiode and stored in the sense node capacitor, before amplification and readout occurs.

However, during the fabrication process, defects may arise in the silicon dioxide layer or may get contaminated with impurities that can trap charge carriers (such as electrons) which will either recombine and be lost, or de-trapped by the defects \citep{le2019leakage}. RTS arises when trapped charge carriers near the source follower channel discretely shift its gain, causing the output voltage to fluctuate between distinct levels. Statistically, RTS is represented as a triple (or more) Gaussian distribution, in which the height of the left/right peaks measures how frequently the event occurs \citep{5410825}. When the images are displayed, the pixel which suffers from RTS, blinks through out a sequence of frames. Other examples of noisy pixels include static and persistent events such as hot pixels, which exhibit high dark current compared to normal pixels making them appear brighter than their surroundings \citep{liu2023effects}. Additional cases include unresponsive/black pixels that fail to efficiently collect or transfer charge.

To address the issue with these spurious pixels, the Marana camera uses a spurious pixel filtering algorithm processed by the FPGA in real-time. The FPGA identifies spurious pixels based on  a selection algorithm which filters pixels whose values exceed the average value of their neighbouring pixels by a certain percentage. The algorithm corrects these pixels by replacing them with the average value of the non-spurious neighbouring pixels within the 3 $\times$ 3 matrix. The spurious pixel filtering can be turned on or off via the Andor SDK3 software. The spurious pixel filter operates near the bias level and uses a predefined threshold to prevent over-correction. It only targets pixels that significantly deviate from the average bias level. As a result, it is expected not to remove actual objects captured by the camera, such as faint stars.

In this paper, we operate the Marana camera with the spurious pixel filter enabled, excluding it only from the read noise analysis to avoid skewing the true read noise distribution.

\section{Laboratory Setup}
\label{sec:equipment}

\begin{figure}
\includegraphics[width=\columnwidth]{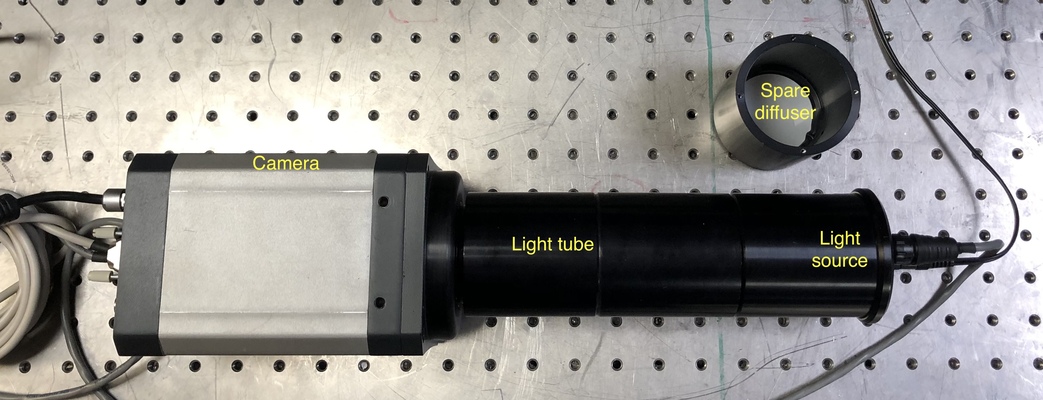}
\caption{Optical bench setup for the Marana camera. The black cylindrical component is the light tube, which houses the light source with a diffuser mounted at the opposite end. A spare diffuser is also visible in the setup.}
\label{fig:camera_bench}
\end{figure}

We tested the Marana camera in the laboratory at the University of Warwick. We employed a DC powered light rig equipped with a 9-volt amber 8-LED light array from OSRAM Opto Semiconductors, which provides consistent illumination peaking at 615\,nm. The cylindrical-shaped rig contained the light source at the one end, and utilized a 100\,mm opal-shaped diffuser to ensure flat illumination. The cylinder rig is at an optimal size to maintain uniform illumination across the sensor area. The diffuser reduces any artefacts that may be caused by small window defects. We covered the internal housing with shielding anti-reflective paper to minimize scattering light, that would normally increase the non-uniformity of the light at the sensor plane. The light tube was securely attached to the front plate of the camera to prevent any stray light from entering the camera as shown in \autoref{fig:camera_bench}. Throughout our testing we kept the light source constant and adjusted the exposure time when needed to vary the illumination for an image frame.

\begin{figure}
\includegraphics[width=\columnwidth]{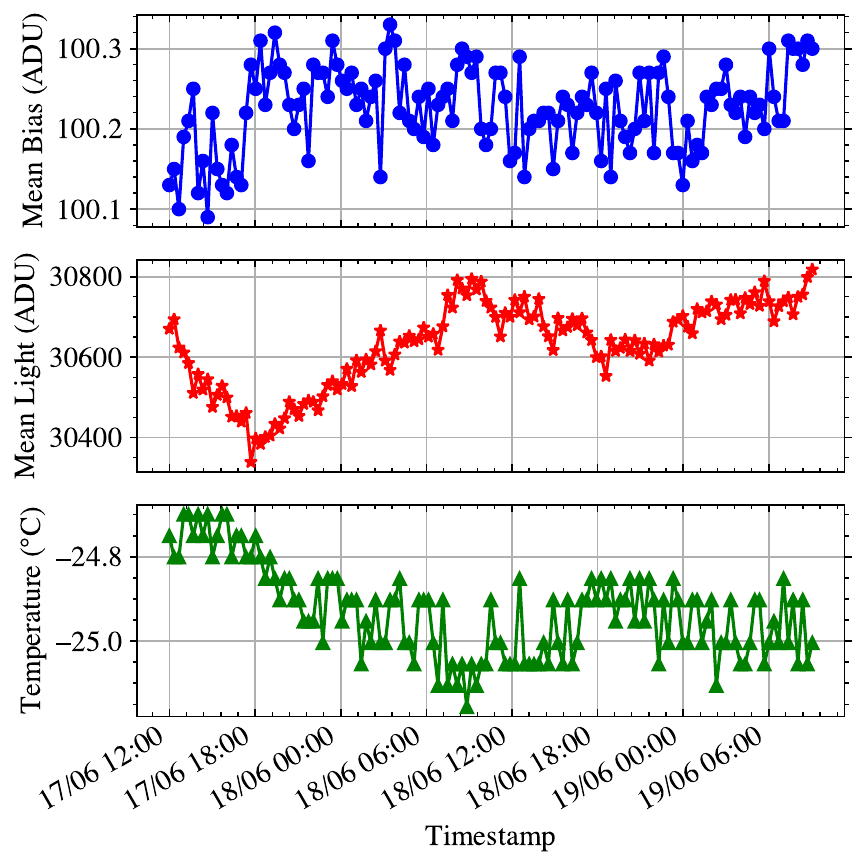}
\caption{Bias level, light source signal and temperature for the Marana camera over time. \textbf{Top:} Average bias level value of a series of images at minimum exposure time and at dark environment. \textbf{Middle:} Average light source signal of a series images at a fixed exposure time. \textbf{Bottom:} Temperature values measured from the camera.}
\label{fig:stability_-25}
\end{figure}

To operate the Marana camera, we used the Andor SDK3 software installed on an Intel NUC 11 Extreme Kit Core mini computer equipped with an i9 CPU, 32\,GB RAM and 4\,TB SSD running the \texttt{Rocky~9} Linux operating system. We controlled the camera using the Andor SDK Version 1.24.0 python wrapper. We integrated the functions of the light rig into the camera control system to directly communicate with the light tube during image acquisition. We configured parameters, including the selection of the Marana camera modes, and the sensor target temperature. To study the stability of the light source signal and the camera sensor temperature, we acquired a set of measurements at 20 minutes interval over the course of 2 days and at a sensor target temperature of -25$^\circ$C. Our results are shown in the middle and lower panel of \autoref{fig:stability_-25}. We monitor the light source stability by acquiring a frame at a fixed exposure that represents the half saturation of the HDR mode ($\sim$30000 analog-to-digital units or ADU hereafter). Similar data are shown in \autoref{fig:stability_15} at 15$^\circ$C.

Certain of the on-board correction algorithms were activated or deactivated to investigate raw sensor performance. These algorithms are the Anti-glow correction (refer to Section~\ref{sec:dark_structure}) and the spurious noise filter (refer to Section~\ref{sec:bias-results}). All results presented in this paper were taken with the sensor cooled at -25$^\circ$C unless explicitly mentioned.

Wherever possible, we adhered to the standard set by the European Machine Vision Association (EMVA), specifically the EMVA-1288\footnote{\url{https://www.emva.org/wp-content/uploads/EMVA1288General_4.0Release.pdf}} latest version (Version 4.0), for Measurement and Presentation of Specifications for Machine Vision Sensors and Cameras, as outlined in \citet{jahne2010emva}.

\section{Methods}
\label{methods}

\subsection{Photon Transfer Curve}
\label{sec:ptc-method}
The Photon Transfer Curve (PTC) was introduced in \citet{1985SPIE..570....7J} and extensively described in \citet{2007SPIE.6690E..03J}.  The PTC is used to extract valuable information on the performance of the camera, such as the conversion gain factor (hereafter gain) and the Full Well Capacity (FWC). The gain value is used to convert pixel values from ADU to the corresponding electrons. We employed this by recording the temporal noise or variance at different signal levels starting from dark conditions until saturation is reached. In order to produce these different signal levels, we used a constant light illumination level and varied the exposure time. \citep{stefanov2022cmos}. 

We record two consecutive frames per exposure. To optimize computational efficiency and acquisition speed, we extracted a 512~$\times$~512 central region from the full-frame images and used an increasing exposure time step as the exposure time increases. Thus, more data points are being captured at lower signal levels. We first estimate the maximum exposure to achieve saturation of the sensor and then we set the exposure time steps. In total, 160 pairs of images were recorded. The exposure time steps are summarised in \autoref{tab:exposure_steps}. Bias frames were acquired to correct for DSNU. The brightness of our light source was kept at a high level to avoid long exposures which could lead to an increase of the noise due to glow and dark current. Given the short acquisition times at various illumination levels, the dark current was negligible; therefore, to further minimize processing time, we did not capture dark frames for each exposure to subtract dark current noise. To distinguish temporal noise (photon shot noise) from spatial noise (FPN), we implemented a technique involving the subtraction of consecutive frames with identical exposures, as outlined in \citep{1985SPIE..570....7J}.  

\subsection{Bias}
\label{sec:bias-method}
Bias is an arbitrary voltage added as an offset to each pixel's value. This is done in order to avoid negative signal values during the readout process which may arise due to random fluctuations of the pixel values \citep{2000hccd.book.....H}. To examine the bias level of the Marana camera, we conducted tests in a dark environment with a cap placed over the camera body to ensure that no light reached the detector. The Dark Signal Non-Uniformity (DSNU) is a measure of the spatial noise distribution on a per-pixel basis across the image array. The DSNU is a result of the column level amplifier structure of the CMOS sensor and pixel defects created during the fabrication of the sensor \citep{stefanov2022cmos}. In this work, we adopt the DSNU parameter from the EMVA Standards \citet{jahne2010emva} as the pattern noise that only depends on bias levels, to remove any contribution that arises due to dark current noise build up at longer exposures. We choose that way in order to avoid other structures produced on dark images such as glow (see Section~\ref{sec:glow}). The DSNU represents the fixed-pattern noise in the bias level, which we typically removed during the astronomical image reduction process and does not affect the final science image. As such, DSNU is generally not considered critical for astronomical applications. However, for the sake of completeness, we include this measurement and report the values, as it may be relevant in other scientific imaging fields.

We defined the temporal variation of pixels in a stack of bias images to be the "readout noise" \citep{fowler2011single}. The readout noise is a combination of all the noise sources that are not dependent on signal or exposure time. For instance, source follower noise which is a combination of thermal noise, the frequency dependent flicker noise and RTS noise. Additionally, the sense-node reset noise, which arises from thermal \citep{johnson1928thermal} voltage fluctuations when the sense node is reset \citep{Janesick2007Photonlambda}; systematic noise generated from the on-board electronics, circuits and power supplies \citep{Janesick2007Photonlambda} and quantisation noise which arises from the conversion of analog to digital values \citep{jahne2010emva}.

In order to isolate the temporal noise from spatial noise, we studied each pixel individually through a stack of images. For DSNU and readout noise tests, the exposure time is set to "minimum", which is $1.52 \times 10^{-5}$\,s for the FFR mode and $9.48 \times 10^{-6}$\,s for the HDR mode. We captured a stack of 100 images in this mode. 

\subsection{Row and Column banding}
\label{sec:row_column-method}
Row pattern, also known as row banding, is a characteristic observed with some CMOS cameras. It is mostly noticeable at very low light levels or at bias level and contributes to the overall read noise of the camera \citep{2014SPIE.9154E..2IW, 2019AN....340..638K, shao2024impact}. The effect is mainly produced from the interference of the camera electronic components and the sensor readout circuits \citep{geurts201798db, 9036882} and, typically, the row patterns are much stronger than their column counterparts \citep{shao2024impact}. Higher frame rates on CMOS cameras may lead to an increase in sensor power consumption, which in turn could lead to higher power supply fluctuations and voltage dips. These electrical fluctuations may manifest in the output images as row pattern noise \citep{geurts201798db}. Other studies have shown that temporal row banding may affect the accuracy of photometry when acquiring short exposure observations \citep{shao2024impact}.

We follow a similar approach as in Section~\ref{sec:bias-method} to study the temporal row-wise and column-wise noise of the Marana camera and report on their contribution to the total read noise of the system.

Advanced methods have been proposed to minimize the row pattern noise such as adding a decoupling capacitor to stabilise the voltage from the row select transistor \citep{CHEN20111265}. For sensors with additional masked pixels on the periphery, these pixels can be used to subtract the offset and eliminate the row pattern on a per-row basis \citep{mo2011method, shao2024impact}.

\subsection{Dark}
\label{sec:dark-method}
The dark current is a signal generated in the absence of light \citep{stefanov2022cmos}. Some electrons are excited from the valence band to the conduction band due to thermal effects within the silicon in the same way as electrons are excited from incident light. Additionally, dark current
is generated by electron leakage from the photodiode or defects in the semiconductor surface, between the silicon and silicon dioxide \citep[e.g.][]{PhysRev.87.835, Aberle1992ImpactOI}. 
%Under the Shockley-Read-Hall (SRH) effect \citep{PhysRev.87.835}, such defects will create electrically active areas by trapping and emitting electrons \citep{Aberle1992ImpactOI}. 
The dark current produced electrons follow the Poisson distribution, and therefore lead to dark shot noise. The dark current increases linearly with exposure time and can be minimized by cooling the sensor \citep{METSELAAR1984320, Widenhorn}. 

We employed tests to measure the dark current in a dark room with the cap on to prevent illumination from background light. The integration time is adjusted to vary from 1 - 10\,s with step of 1\,s. We did not subtract the bias level from the dark signal data. The slope of the dark signal as a function of time yields the dark current and the intercept the bias level. The same method was used at different temperatures to investigate the dependence of the dark current with temperature.

\subsection{Photo Response Non-Uniformity}
\label{sec:prnu}
The Photo Response Non-Uniformity (PRNU) characterizes the variability in pixel responsiveness to uniform light levels. Differences in sensor manufacturing processes can lead to variations in pixel sensitivities, causing some pixels to capture more or less electrons than others \citep{stefanov2022cmos}. This results in a spatial structure in the frame, where each pixel exhibits a unique sensitivity profile. Different source followers can produce different conversion gain values for each pixel, that can also significantly contribute to higher PRNU \citep{jain2016characterization}.

We characterise the spatial noise under light conditions as Fixed Pattern Noise (FPN), which manifests as pixel-to-pixel response variations across the image array. Specifically, we investigate the PRNU, which is defined as the ratio of the FPN to the mean level of the signal. It represents the proportionality constant that relates the signal level to the FPN amplitude \citep{2007SPIE.6690E..03J}. 

To quantify the PRNU, we recorded 100 illuminated and bias subtracted images with constant light exposure at a level of approximately half the digitised saturation. We suppressed the photon shot noise components by averaging the 100 images to assess the response of individual pixels across the image \citep{ye2018cmos}.
 
A potential source of non-uniformity which is unrelated to the sensor itself is the appearance of dust or other debris on the optical window of the camera. To avoid this complication, we ensured the camera window was cleaned before testing using methanol and lens cleaning tissues. During testing, the camera was kept affixed to the light-box, as described in Section~\ref{sec:equipment}, to prevent any new dust accumulating on the camera window. 

\subsection{Linearity Testing}
\label{sec:linear-method}
Linearity defines the relationship between the measured signal and the illumination level or integration time \citep{stefanov2022cmos}. It is characterised by a linear correlation between the signal and exposure time. When treating the camera as a black box—where incident photons enter and a digitised signal is output—linearity can be evaluated by either keeping the integration time constant while varying the light source intensity or maintaining a fixed light source while adjusting the integration time.

Measuring linearity in CMOS image sensors, particularly when employing a dual amplifier structure from the HG and LG channels, is crucial due to the presence of various effects that can impact overall linearity. In CMOS sensors with active pixel architecture, additional complications arise during the multiple process stages of the signal processing chain. This includes the linearity of the voltage amplification by the source follower transistor, as well as the linearity of the Analog-to-Digital conversion into ADU \citep{wang2017linearity}. Each stage in this processing chain has the potential to introduce non-linearities. To mitigate these effects, fabrication techniques for CMOS sensors have been proposed \citep{wang2017linearity, wang2018development}, effectively improving linearity performance. Additionally, studies have identified a correlation between non-linearity and both the modulation transfer function and pixel noise \citep{li2016study}. Furthermore, compensation techniques and on-chip calibration methods for addressing non-linearities have been reported \citep{wang2004capacitance, wang2018development}.

We explored how the signal changes relative to exposure time. We used a stable light source fixed while varying the exposure. For this investigation we utilised the same dataset as used in the PTC (see Section~\ref{sec:ptc-method}). 

\subsection{Quantum Efficiency and Window Transmittance}
\label{sec:QE_method}
\begin{figure*}
\includegraphics[width=\textwidth]{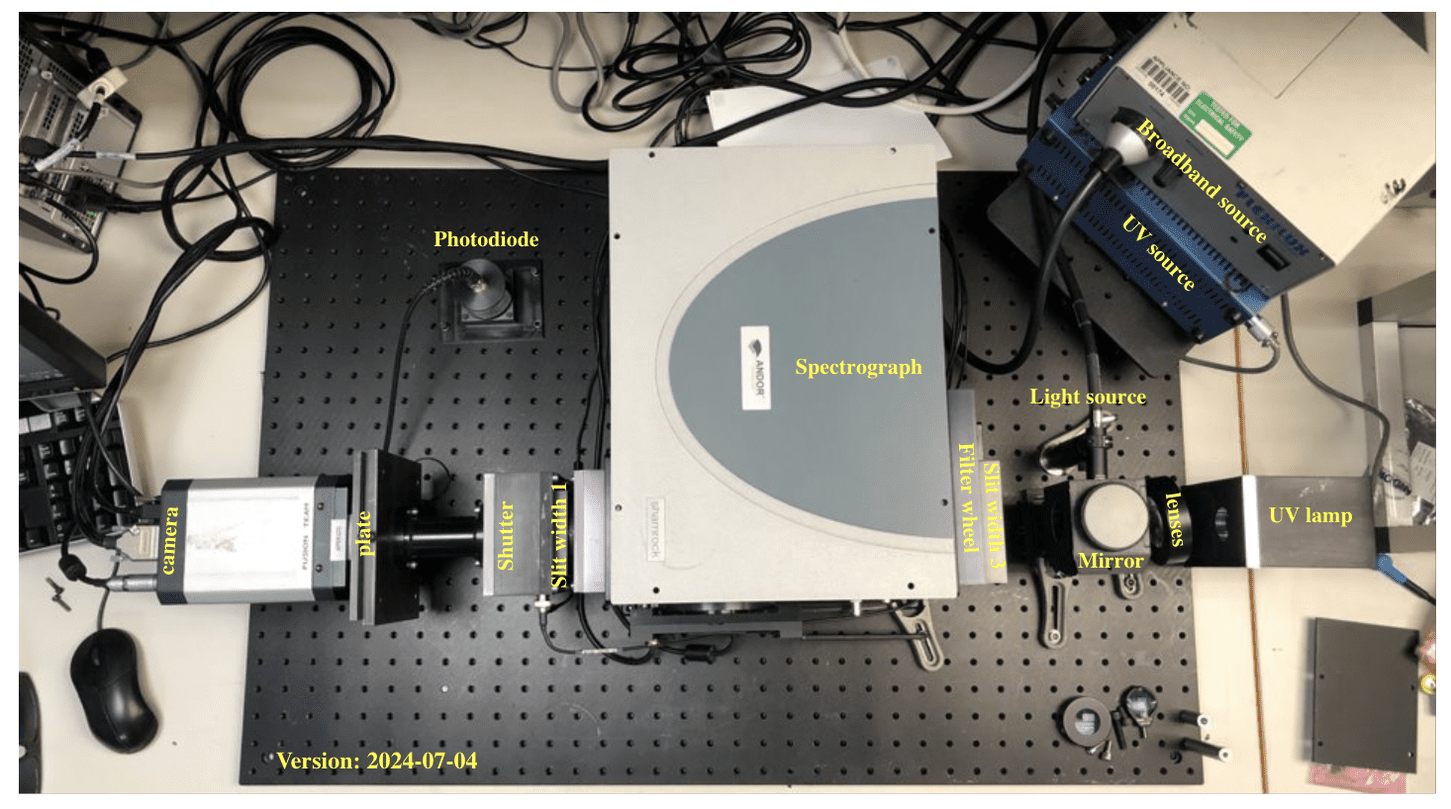}
\caption{The optical bench of the QE measurement of the Marana CMOS camera, including the UV light bulb. Each instrument is labelled with yellow coloured text.}
\label{fig:Real_Schematic}
\end{figure*}

Many different approaches have been reported in the literature for Quantum Efficiency (QE) measurement \citep{2014MeScT..25a5502S, Gill_2022, Shao_2024, 2025arXiv250200101L}. In this work, we set out the methodology and results for our QE testing of the Marana CMOS camera, following the approach of \cite{2025arXiv250200101L}. 
The QE experiment was conducted at Andor's laboratories. A labelled photograph of the setup is shown in \autoref{fig:Real_Schematic}. 

A halogen bulb is used in order to provide constant illumination across the visible spectrum. The light received by the photodiode is converted to electric current and measured with the use of a Keithley picoammeter. To make a measurement at a desired wavelength, Andor's Shamrock SR-303i spectrograph is used as a monochromator. The data acquisition from both the photodiode and camera has been automated using a custom made code. This code establishes communication with the spectrograph in order to set the slits widths, filters, and grating angle and also allows data to be retrieved from the picoammeter through an RS232 protocol.

The wavelength ranges from 400 to 1020 nm in steps of 5\,nm. The width for both entrance and exit slits was set to 1000\,$\mu$m and was found to be sufficiently wide for enough light to come through while short enough to only allow a small range of wavelengths at the output. By setting the slit width to 1000\,$\mu$m, the camera exposure time was limited to a range of 0.001 to 0.01 seconds to avoid saturation. We used a combination of filters that are the most sensitive at each wavelength range to eliminate stray light entering the monochromator. First, we measure the photodiode current across the desired wavelength range by mounting it at the exit slit of the monochromator. The photodiode is then replaced by the Marana CMOS camera which is configured at a fixed exposure of 1\,ms and the measurement is repeated.

The camera front window consists of a quartz substrate and is covered with an anti-reflection coating which has been specifically chosen to match the sensor's quantum efficiency. This means that the total camera quantum efficiency will be the product of the sensor quantum efficiency and the optical window transmission. Thus, in order to isolate the sensor quantum efficiency, we measured the transmittance of an optical window that is identical to the one used in the Marana camera, with the same glass type, thickness, and anti-reflective coating, using the experimental setup described above. In short, the transmission window is positioned between the entrance slit of the monochromator and the optical fibre connected to the light source with the photodiode mounted at the exit slit in order to record the photodiode current at various wavelengths. The same measurement is then repeated after the window has been removed.

\section{Results, Analysis and Discussion}
\label{results}

\subsection{Photon Transfer Curve Results}
\label{ptc}
To assess the characteristics of the Marana camera such as the gain and the FWC, we used the PTC technique as described extensively in \citet{1985SPIE..570....7J}. We take two consecutive frames (k=A and k=B) which are used to measure the average signal over all the pixels using the following equation for each (adopted from \citet{bohndiek2008comparison}):

\begin{equation}
    \mu_y[k] = \frac{1}{MN}\sum_{m=0}^{M-1}\sum_{n=0}^{N-1}y[k][m][n], \\ (k = A, B)
\end{equation}

and the average signal from the two frames is calculated as:
\begin{equation}
    \mu_{\mathrm{A,B}} = \frac{1}{2} ( \mu_y[A]  + \mu_y[B])
\end{equation}
   
The temporal variance is not calculated directly from the averaged set. For illuminated images, the main noise sources are the photon shot noise and the fixed pattern noise. The read noise is signal-independent and therefore considered constant. The maximum dark current contribution for the exposure ranges used to construct the PTC curves are 0.022 and 0.26 electrons for the FFR and the HDR mode, respectively. Therefore the read noise and the dark current contribution are considered negligible.

While the FPN is the spatial variance of the system, the photon shot noise is the temporal and their expression is given as:

\begin{equation}
       \sigma^2_{y} + s^2_{y} = \frac{1}{NM}\sum_{m=0}^{M-1} \sum_{n=0}^{N-1} (y[k][m][n] - {\mu}_y[k])^2, \\ (k = A, B)
\end{equation}

where the FPN denoted as $s^2_{y}$ and the photon shot noise as $\sigma^2_{y}$. To eliminate the FPN, one has to subtract two consecutive frames at the same exposure \citep{2007SPIE.6690E..03J}. Subtracting two consecutive frames removes the spatial variance associated with the FPN and leaves only the temporal noise component as in the following: 

\begin{equation}
\label{eq:variance}
    \sigma_{y}^2 = \frac{1}{2NM} \sum_{m=0}^{M-1} \sum_{n=0}^{N-1}(y[A][m][n] - \mu_y[A]) - (y[B][m][n] - \mu_y[B]))^2
\end{equation}

However, when a set of two frames are added or subtracted, the random variance component of the resulted frame increases by a factor of 2. Therefore, \autoref{eq:variance} is divided by this factor to extract the true pixel temporal variance \citep{2007SPIE.6690E..03J}. 

\begin{figure*}
\includegraphics[width=\textwidth]{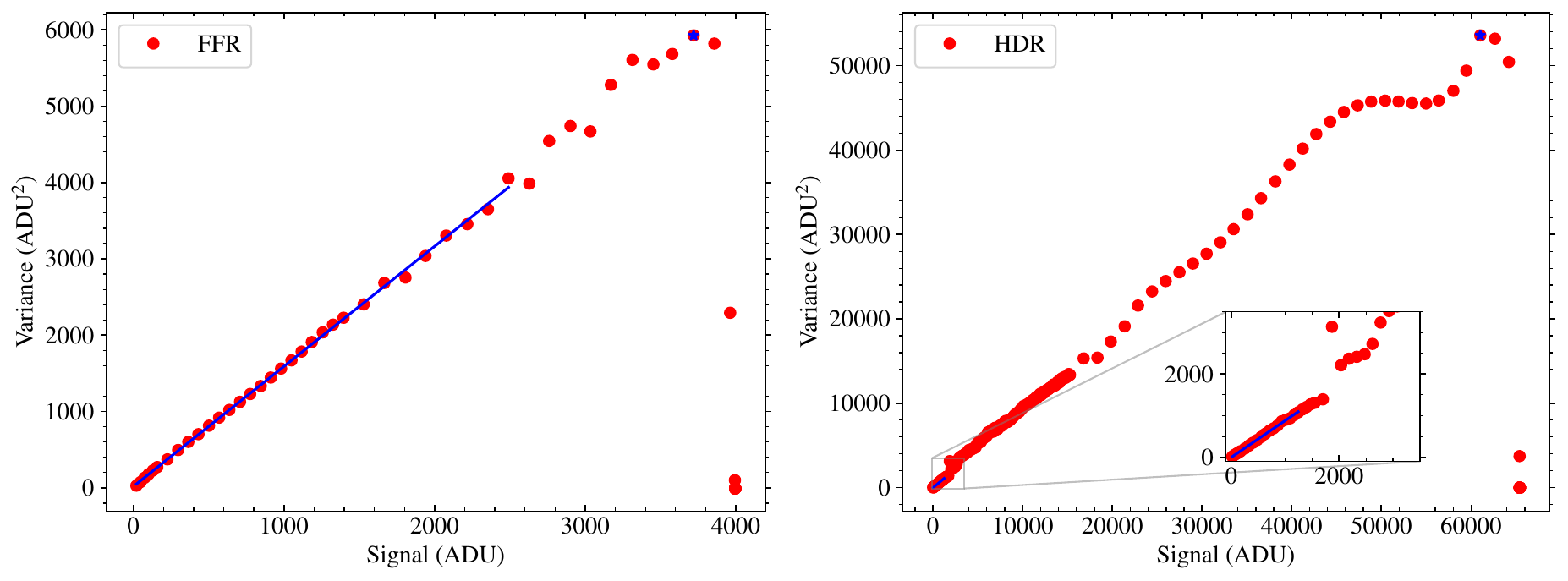}
\caption{Photon transfer curves for the Marana camera at -25 $^{\circ}$C. \textbf{Left:} The FFR readout mode. \textbf{Right:} The HDR readout mode. The solid blue line is a best fit linear model. The data point with the highest variance is marked with a blue star, and indicates the FWC value. A zoom-in panel illustrates the transition region.}
\label{fig:ptc}
\end{figure*}

The PTC curves for both FFR and HDR modes are shown in \autoref{fig:ptc}. The y-axis indicates the photon shot variance which has been isolated following the method described above. A linear regression is applied in the range of 0-70\,\% from the saturation, in order to extract the slope of the system which represents the inverse gain G (ADU/e$^-$). This is expressed as:

\begin{equation}
\label{eq:gain}
    \sigma^2_y = \frac{\mu_{\mathrm{A, B}}}{G} + \sigma^2_{\mathrm{RN}} 
\end{equation}

Where $\sigma^2_y $ is the photon shot variance, $\sigma^2_{\mathrm{RN}}$ is the read noise variance, $\mu_{\mathrm{A, B}}$ is the mean signal and $1/G$ is the gain which is measured to be $1/G=0.632\pm0.009$\,e$^-$/ADU for the FFR mode and $1/G=1.131\pm0.008$\,e$^-$/ADU for the HDR mode. We adopted these gain values for the Marana camera and used them to convert our results to electron units (i.e. read noise and dark current).

The FWC can be extracted from the x-axis for the maximum system variance. The variance decreases while the pixels start becoming saturated, since pixel values reach an identical state, indicative of sense node saturation. The FWC is measured to be 3720\,ADU for the FFR and 61031\,ADU for the HDR mode. Using the gain values the FWC is measured to be 2351\,$e^-$ for the FFR and 69026\,$e^-$ for the HDR mode. This was in a good agreement with the FWC in the performance sheet of the camera (see \autoref{tab:specs}). It is worth noting that we do not calculate the FWC based on the linearity curves, given that the graph only levels off once it reaches the digitisation saturation.

The HDR mode shows a clear transition region at approximately 1800\,ADU, which can be seen in the right and zoomed panel of \autoref{fig:ptc}. The transition is also evident in previous studies reported in \citet{2021RAA....21..268Q} and \citet{2019AN....340..638K}. The transition region under uniform illumination corresponds to the switch from the HG to the LG channel (see Section~\ref{sec:hglg} and \citet{ma2013low, tang2020high} for details). The LG channel has larger read noise ($\sim$40\,$e^-$)\footnote{A separate analysis of the HG and LG channels confirmed this read noise difference; however, since this work focuses on overall HDR and FFR mode performance, those results are not presented here.} compared to the HG channel. Since it is a noise component, it contributes directly to the total variance which is shown as a discontinuity in the PTC for the HDR mode in \autoref{fig:ptc}. When the signal exceeds the HG channel’s capacity, excess charge is transferred to the LG channel. To mitigate the gap caused by amplifier gain differences, the amplifier gain ratio could be adjusted to smaller values, however, this adjustment would lead to either higher read noise (by reducing the HG amplifier gain value) or compromise the maximum FWC by increasing the LG amplifier gain value (see \citet{tang2020high} for details). As a result, linear regression was exclusively applied to low signal values within the range of range 0-70\,\% for the HG channel, precisely before entering the transition region.

Interestingly, the pixel values on the transition point ($\sim$1800\,ADU) may be processed by both HG and LG channels. If a pixel is near the transition threshold an increase on its output value can be observed throughout a series for the same exposure and illumination levels. This is because the shot noise sometimes drives the pixel to above or below the transition threshold value. Importantly, this dynamic channel assignment has repercussions when one subtracts frames consecutively, which may lead to higher variance value as shown with an additional spike in the transition in the zoom-in panel of \autoref{fig:ptc}.

To investigate this, we acquired a set of 150 sub frames (460 $\times$ 250) under constant illumination and exposure settings, while applying a gradient in the illumination in the range of 1200-2000\,ADU. This range was chosen to isolate pixels operating in distinct amplification modes: at lower signals (<1800\,ADU), only the HG amplifier is active, while at higher signals (>1800\,ADU), the LG amplifier is used exclusively. In the transition region around 1800\,ADU, however, pixel values may be amplified by either the HG or LG channel, depending on the specific pixel behaviour.

We measured the temporal noise of each pixel across the frames. Pixels undergoing a transition between channels are expected to exhibit noise levels exceeding the combined shot and read noise at each signal level. For example, a pixel read out at 1400\,ADU using the HG channel should have an expected noise of approximately 37\,ADU, while a pixel at 2000\,ADU in the LG channel should have expected noise around 60\,ADU. Any pixel with noise exceeding these expected values likely experiences additional noise due to the HG–LG channel transition.

\begin{figure}
\includegraphics[width=\columnwidth]{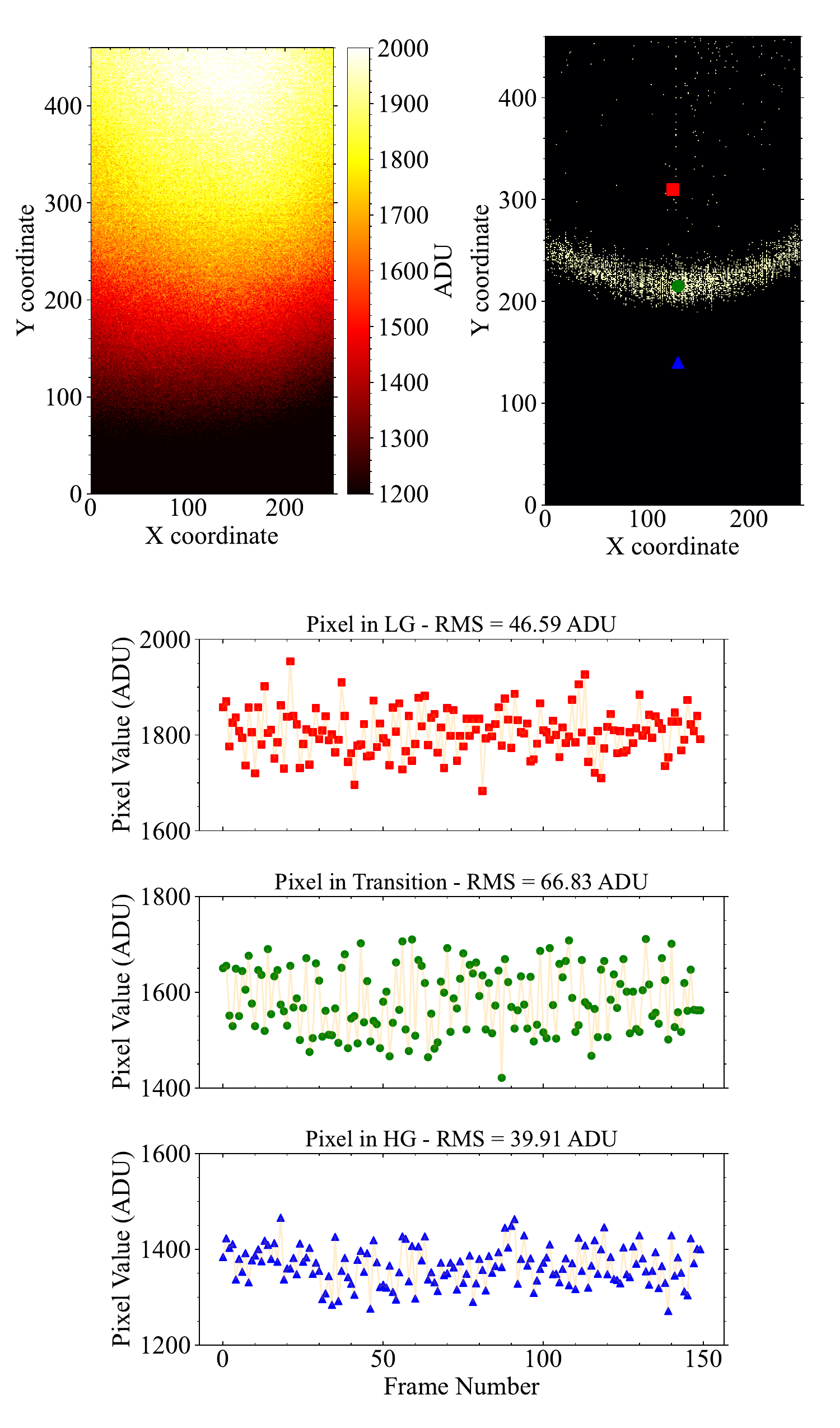}
\caption{\textbf{Top left:} An example sub-frame showing the illumination difference across the LG and HG transition region. \textbf{Top right:} the binary image is highlighting pixels that exceed the threshold noise value. Also shown are three example pixels processed with the LG, HG channel and in transition, indicated with the red square (Y,X = 310, 125), blue triangle (Y,X = 140, 130) and green circle (Y, X = 215, 130) respectively. \textbf{Bottom:} The ADU counts for the three example pixels as a function of frame number, with the RMS calculated for each pixel.}
\label{fig:Binary_Time_Series}
\end{figure}

To analyse this, we applied a noise threshold of 60\,ADU. Pixels exceeding this threshold were flagged, resulting in a binary image, as shown in the top-right panel of \autoref{fig:Binary_Time_Series}, where only the high noise pixels are highlighted. Additionally, example pixels processed with the HG channel, LG channel and the transition region were selected and plotted as a function of frame number in \autoref{fig:Binary_Time_Series}. 

The dual amplifier architecture, which constructs an HDR image from HG and LG channel may influence our comparison stars. Specifically, dim stars with low counts will be processed using the HG channel, while bright stars will be processed using the LG channel. Additionally, since the sky background remains considerably low, it will be read out using only the HG channel. 

The PTC of the HDR mode exhibits significant non-linear artifacts, particularly dips and jumps at signal levels above the transition point and up to saturation. These effects can potentially be attributed to the non-linear conversion gain of individual pixels. Other studies have also shown that charge transfer inefficiency, also referred to as image lag, can contribute to such behaviour \citep{stefanov2022cmos}. This charge spill-back arises from incomplete charge transfer between the photodiode and the sense node, often due to inefficient activation of the transfer gate \citep{Teranishi}. Furthermore, \citet{michelot2016effects} demonstrated that this non-linearity in the PTC can be mitigated by maintaining a high reset voltage while keeping the transfer gate low. Image lag can affect photometry results, as residual charge on the photodiode may be lost or accumulate onto the next image during exposures.

\subsection{Bias level, Read Noise and DSNU}
\label{sec:bias-results}
We calculate the mean value ($\mu_{y}$) of a bias frame (k) by averaging over all of the pixels as:
\begin{equation}
\label{eq:average_bias}
 \mu_{y}[k] = \frac{1}{NM} \sum_{m=0}^{M-1} \sum_{n=0}^{N-1} y[k][m][n].   
\end{equation}
We follow the notation of the EMVA-1288 standards version 4.0 \citet{jahne2010emva}, where $y[m][n]$ is the value of the pixel in ADU for the $m$th row and $n$th column. $M$ and $N$ are the total number of active pixels in the rows and columns of the detector, which in the case of the Marana camera is M=2048 and N=2048.

\begin{figure*}
\includegraphics[width=\textwidth]{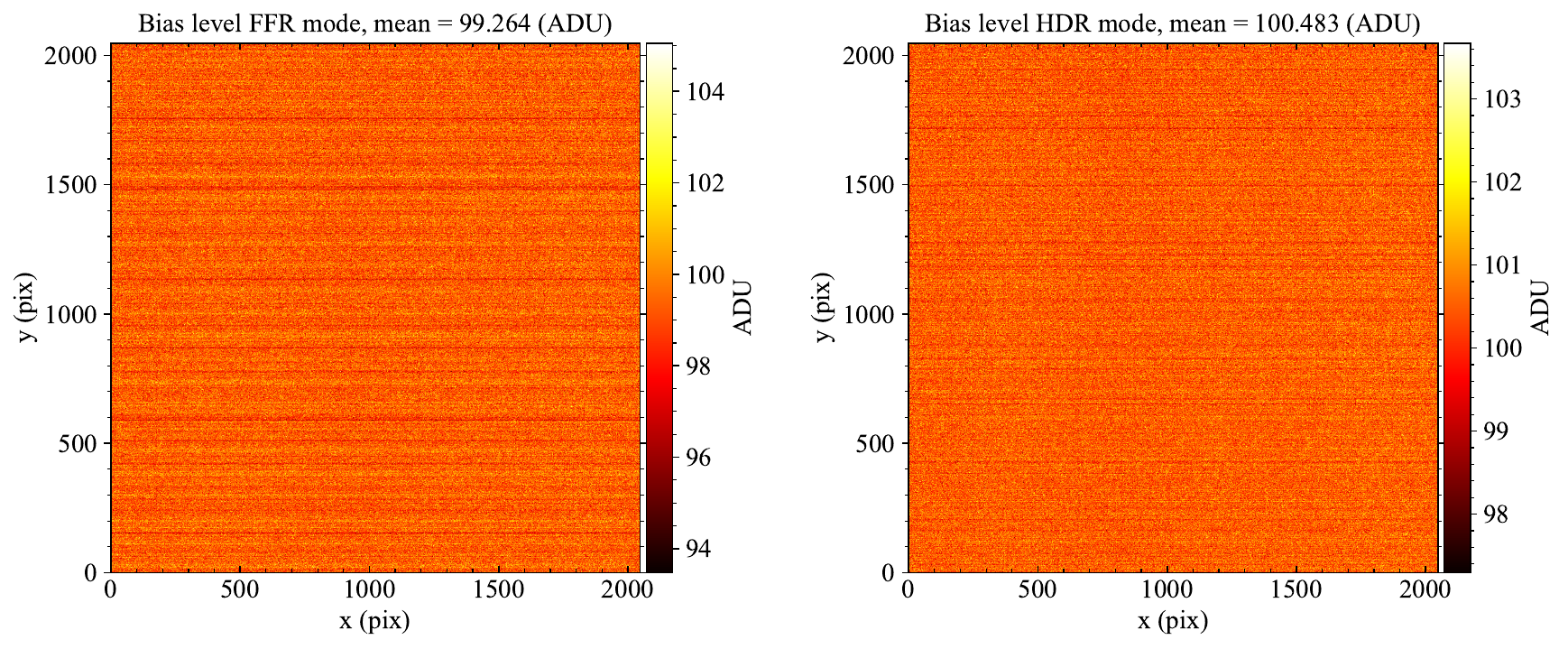}
\caption{Bias frames from Marana camera recorded at minimum exposure for the FFR (\textbf{left}) and the HDR (\textbf{right}) modes. The camera is operating at temperature of -25$^\circ$\,C.}
\label{fig:bias_frames}
\end{figure*}

We find the mean value for the bias level of the Marana camera to be 62.74\,e$^-$ (99.26\,ADU) with a standard deviation of 1.81\,e$^-$ in FFR mode, and 113.65\,e$^-$ (100.48\,ADU) with a standard deviation of 1.78\,e$^-$ in HDR mode. No large scale structures are visible in the bias frames - see \autoref{fig:bias_frames}. Some low level (1<\,e$^-$) horizontal banding can be seen in both modes. The patterns are slightly more prominent in the FFR mode. The banding is random in nature and cannot be corrected out by means of FPN subtraction \citep{2014SPIE.9154E..2IW, Shao_2024}. The bias values are in agreement with the bias values in \cite{2019AN....340..638K}. However, \cite{2021RAA....21..268Q} reported the bias level to be 141\,ADU for the HDR and 207\,ADU for the FFR mode.

To investigate the pixel-level \textit{temporal} noise in bias frames, we collect a set of K sequential bias images. We organize the bias images into a 3-D array, with the $m$ and $n$ dimensions representing pixel row and column positions on the detector and $k$ representing the image index. This technique ensures that we can study each pixel individually avoiding the dark fixed pattern noise contribution. For each of the 4,194,304 (M=2048 and N=2048) individual pixels across the frames, we calculate the mean as:
\begin{equation}
 \mu_{m,n} = \frac{1}{K} \sum_{k=0}^{K-1} y[k][m][n],   
\end{equation}

and standard deviation as:

\begin{equation}
\sigma_{m,n} = \sqrt{\frac{1}{K-1}\sum_{k=0}^{K-1}(y[k][m][n] - \mu_{m,n})^2}. \\
\end{equation}

\begin{figure}
\includegraphics[width=\columnwidth]{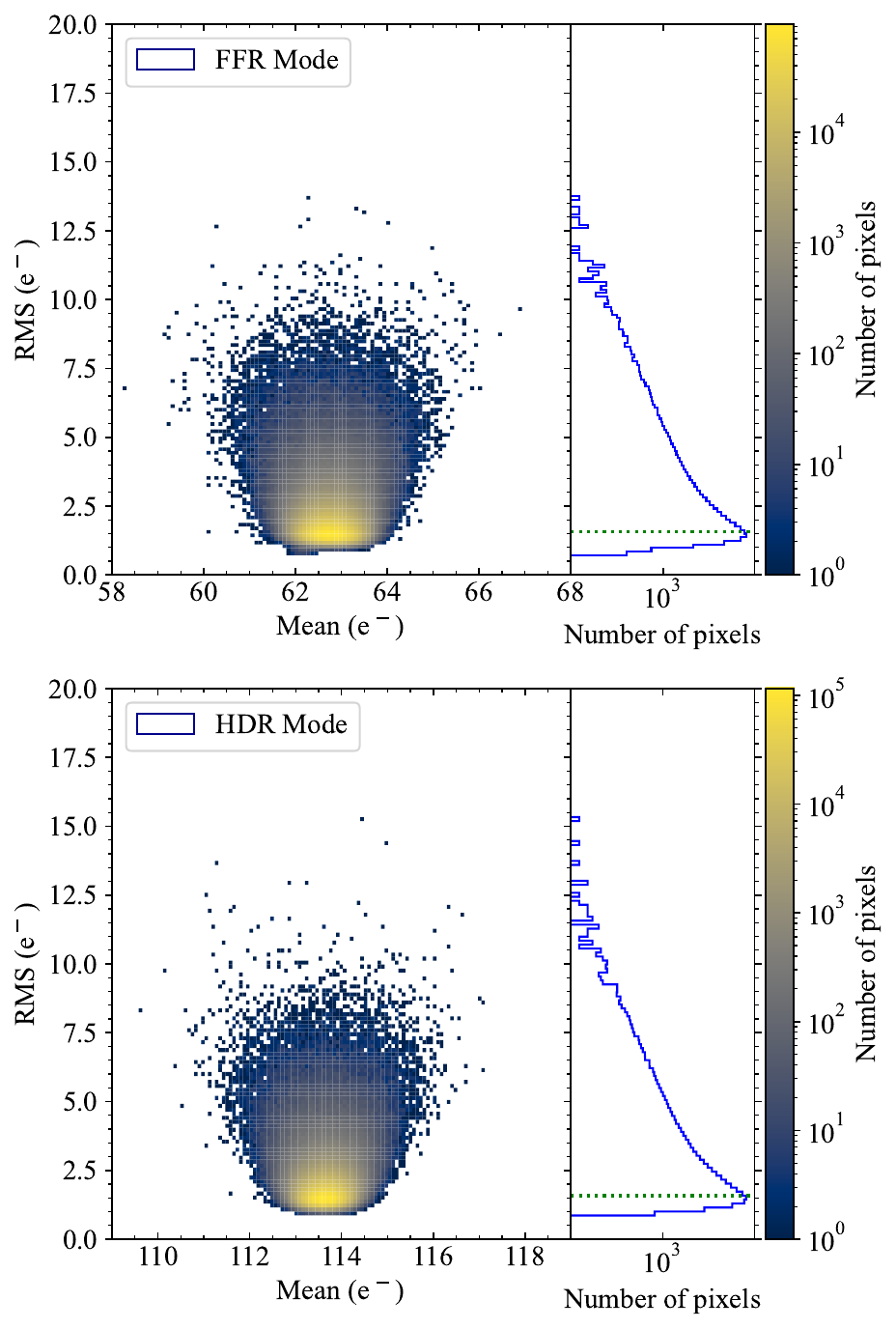}
\caption{Read noise distribution using 100 consecutive bias frames. The 2-D histogram corresponds to temporal noise. The scattered plot shows the mean value of each pixel as measured} from these frames while the colour bar indicates the number of pixels. The logarithmic read noise distribution is shown in the blue histograms, with the median value as a green dotted line. \textbf{Top:} Median read noise distribution for the FFR mode. \textbf{Bottom:} Median read noise distribution for the HDR mode.
\label{fig:rn}
\end{figure}

We carry out this calculation for 100 bias images in the FFR mode and 100 bias images in the HDR mode. The mean values and standard deviations are visualized through the 2-D histograms presented in \autoref{fig:rn}. We calculate the mean read noise for the camera ($\textrm{RN}_{\textrm{mean}}$):
\begin{equation}
\textrm{RN}_{\textrm{mean}} =  \frac{1}{NM} \sum_{m=0}^{M-1} \sum_{n=0}^{N-1} \sigma_{m,n},\\
\end{equation}

and the median read noise is found as the middle value after sorting all the $\sigma_{m,n}$ values. 

The mean, median and RMS read noise for the Marana camera in FFR mode is 1.71\,e$^-$, 1.577\,e$^-$ and 1.763\,e$^-$ respectively. The mean, median and RMS read noise in HDR mode is 1.697\,e$^-$, 1.571\,e$^-$ and 1.785\,e$^-$ respectively. The median read noise values are in good agreement with the performance metrics provided by Andor. The results are very similar between the two modes, since both utilize the HG channel at low illumination levels. Since the camera operates with the LG channel for high signal levels in HDR mode, the read noise after the transition point at 1800\,ADU increases to $\sim$ 40\,e$^-$. Using a similar pixel-wise method to estimate the read noise, our read noise measurement is approximately half of the values reported in \cite{2024SPIE13103E..0RK} and in \cite{2021RAA....21..268Q} with a measured read noise of 3\,e$^-$ and 3.1\,e$^-$ respectively. Moreover, the read noise from the Marana CMOS is lower than comparable CCD cameras such as the iKon-L, which has a read noise of 15\,e$^-$ at 3\,MHz speeds \citep{2018MNRAS.475.4476W}. CCDs can also achieve low read noise, however it comes at the expense of reduced readout speeds.

We note that for both readout modes the distribution of the read noise is non-Gaussian, and shows a very long tail of high read noise values (see histograms in \autoref{fig:rn}). There are two reasons that contribute to the long read noise tail in \autoref{fig:rn}. The first reason is that in the CMOS architecture there are differences in the per-pixel amplifiers and column ADC converters. Therefore each pixel has its own read noise properties, unlike a CCD where pixels are readout through a common amplifier and ADC converter. This can result in a skewed distribution of read noise if some amplifiers or ADC converters are noisier than others. The second reason is the impact of RTS noise, which have also been reported for the Marana in previous work \citep{2019AN....340..638K}, and forces random pixels towards larger read noise values \citep{5410825} and skew distribution on the histograms in \autoref{fig:rn}. 

\begin{figure}
\includegraphics[width=\columnwidth]{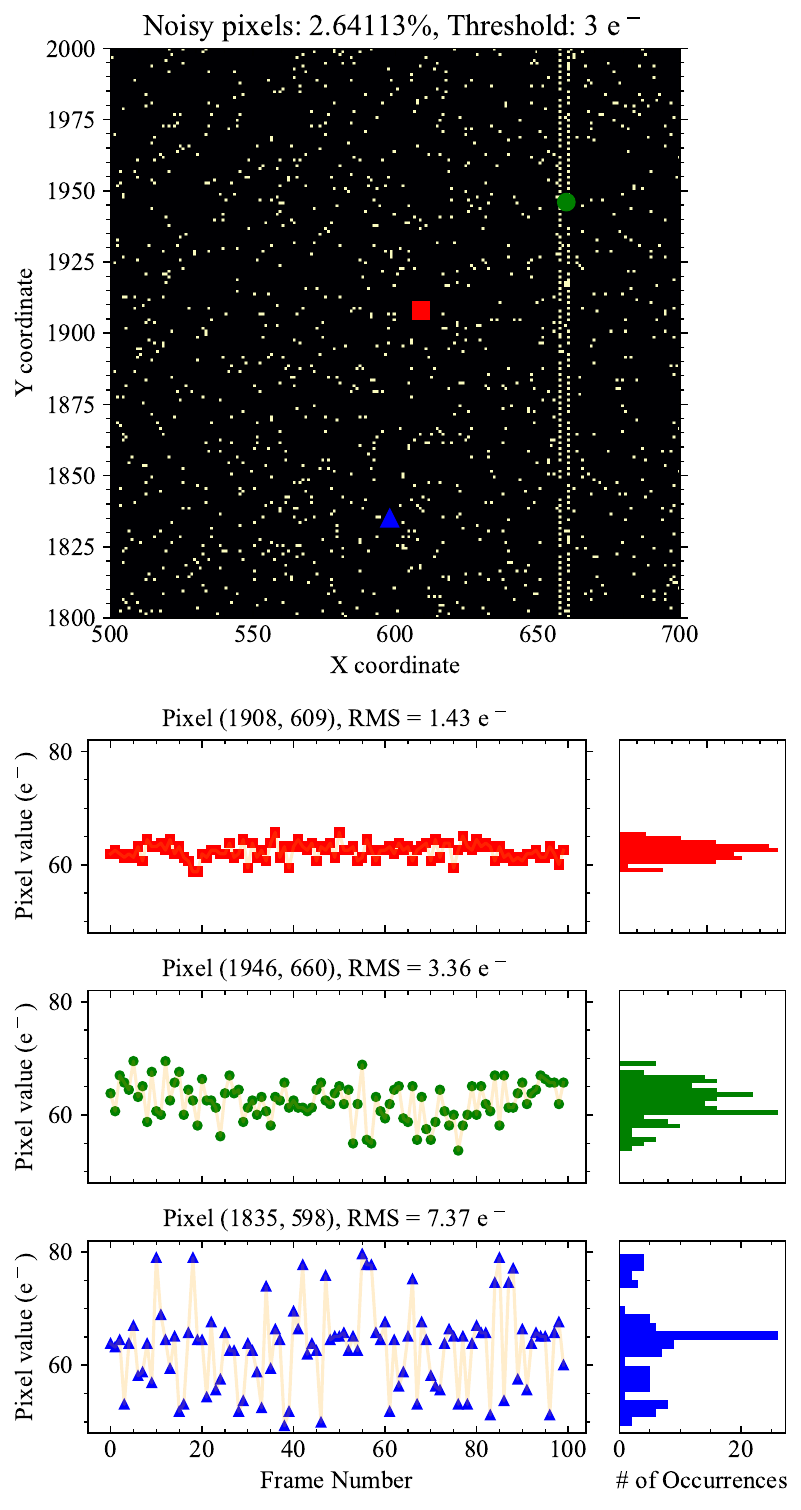}
\caption{An example sub-frame showing the read noise map for the FFR mode. \textbf{Top:} Binary image is highlighting pixels that exceed the threshold read noise value. Also shown three example pixels below the threshold, in a noisy column level amplifier and noisy pixel with RTS behaviour, indicated with the red square (Y,X = 1908, 609), green circle  (Y,X = 1946, 660) and blue triangle  (Y,X = 1835, 598) respectively.\textbf{Bottom:} The e$^-$ values for the three example pixels as a function of frame number, with the RMS calculated for each pixel.}
\label{fig:rts}
\end{figure}

To demonstrate this, we created a binary image as shown in \autoref{fig:rts}, using 3\,e$^-$ read noise threshold. Pixel values that exceed this threshold are considered noisy. There are various pixels that exceed this threshold, in total 2.64\% of the image, most of the spread randomly. However, there are two distinctive column level amplifiers that are noisy. For pixels below the threshold, their bias level at different times are consistent with a Gaussian distribution. In contrast, pixels within the noisy column-level amplifiers show a non-uniform temporal distribution, while pixels outside these columns that exceed the threshold likely display RTS noise, characterized by a multimodal distribution as shown in \autoref{fig:rts}.

To further investigate noisy pixels, we used 100 bias images with and without the corrections in order to investigate the number of the corrected pixels. We found that less than 0.02\,\% of the total pixels on a typical bias frame are flagged as noisy pixels, some of which may also exhibit RTS and are detected by the Spurious Noise Filter on the FFR mode. We did not see any particular difference in the HDR mode. The Spurious Noise Filter operates at very low light signals, approximately to the bias level, and it works by removing noisy pixels refereed to as spikes.

%DSNU
To measure the DSNU, we employ the same set of K bias images for the read noise analysis as in Section~\ref{sec:bias-results}. As per the EMVA-1288 standard, for modern CMOS sensors, pixels, rows and columns non-uniformities should also be included in the non-uniformity analysis. To suppress the temporal noise (i.e. read noise) we average these frames using the equation:

\begin{equation}
\label{eq:DSNU}
    \Bar{y}_{\mathrm{k}} = \frac{1}{K}\sum_{k=0}^{K-1} y[k]
\end{equation}

Where $\Bar{y}_{\mathrm{k}}$ is the averaged frame. Using the mean value $\mu_{y. \mathrm{bias}}$ of the averaged frame as in \autoref{eq:average_bias}, the spatial variance of the image is calculated using the following:

\begin{equation}
s_{\mathrm{y}}^2 = \frac{1}{MN}\sum_{m=0}^{M-1}\sum_{n=0}^{N-1}(\Bar{y}_{\mathrm{k}}[m][n]-\mu_{y. \mathrm{bias}})^2
\end{equation}

This expression, however, includes a contribution from the temporal variance. To isolate the true spatial variance, we subtract the temporal component:

\begin{equation}
s_{\mathrm{y.k}}^2 = s_{\mathrm{y}}^2 - \frac{\sigma_{y}^2}{K}
\end{equation}

where $s_{\mathrm{y.k}}^2$ is the spatial variance values for the overall image. To investigate the row and column non-uniform structure, we have to solve for the columns:

\begin{equation}
    \mu_y[n] = \frac{1}{M}\sum^{M-1}_{m=0}\Bar{y_k}[m][n]
\end{equation}

where $\mu_y[n]$ is the mean value for the columns for the averaged frame. The same approach is done on the rows:

\begin{equation}
    \mu_y[m] = \frac{1}{N}\sum^{N-1}_{n=0}\Bar{y_k}[m][n]
\end{equation}

where $\mu_y[m]$ is the mean value for the rows for the averaged frame. Again, for both rows and columns we have to remove the temporal variance across rows and columns, respectively \citep{jahne2010emva}:

\begin{equation}
    s^2_{y.m} = \frac{1}{M}\sum^{M-1}_{m=0}(\mu_y[m] - \mu_y)^2 - \frac{s^2_{y.pix}}{N} - \frac{\sigma^2_y}{KN}
\end{equation}

\begin{equation}
    s^2_{y.n} = \frac{1}{N}\sum^{M-1}_{m=0}(\mu_y[n] - \mu_y)^2 - \frac{s^2_{y.pix}}{M} - \frac{\sigma^2_y}{KM}
\end{equation}

Where $s^2_{y.m}$, $s^2_{y.n}$ and $s^2_{y.pix}$ are the spatial variances for rows, columns and pixel. The total spatial variance is expressed as:

\begin{equation}
s_{y,k}^2 = s_{y,m}^2 + s_{y,n}^2 + s_{y,\text{pix}}^2
\end{equation}

Finally, by solving the matrix with the above three components \citep{jahne2010emva} the corresponding DSNU values are estimated as:

\begin{equation}
\begin{aligned}
s_{\mathrm{DSNU}} &= \sqrt{s^2_{y.k}} \\
s_{\mathrm{DSNU, row}} &= \sqrt{s^2_{y.m}} \\
s_{\mathrm{DSNU, col}} &= \sqrt{s^2_{y.n}} \\
s_{\mathrm{DSNU, pix}} &= \sqrt{s^2_{y.k} - s^2_{y.m} - s^2_{y.n}} \\
\end{aligned}
\end{equation}

\begin{figure}
\includegraphics[width=\columnwidth]{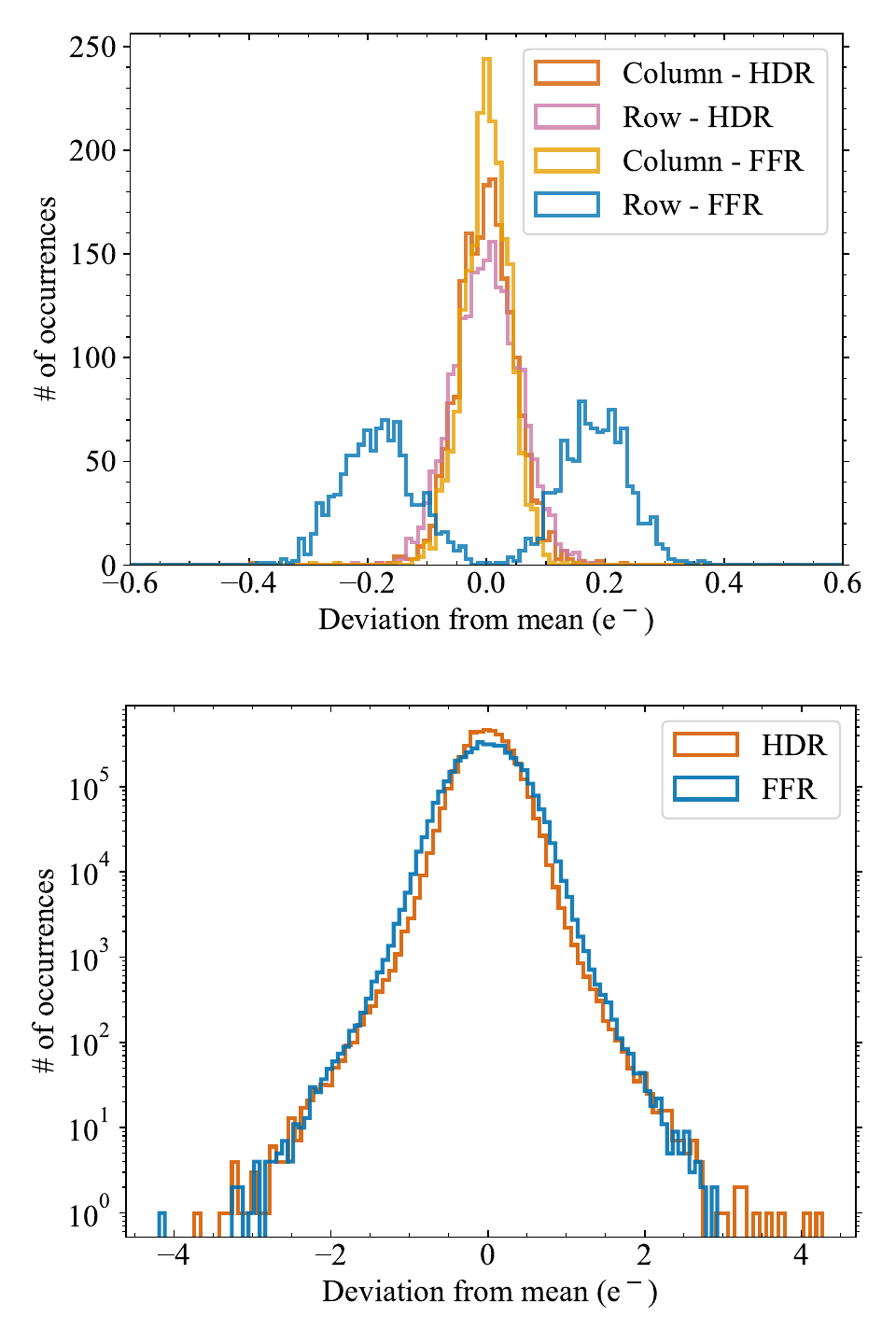}
\caption{Histogram distribution of the mean-subtracted pixel values used to estimate the DSNU for both readout modes. \textbf{Top:} Distribution of the row- and column-averaged pixel values after subtracting the spatial mean level. \textbf{Bottom:} Distribution of the image-averaged pixel values after subtracting the spatial mean level.}
\label{fig:dsnu}
\end{figure}

All DSNU values are reported in electrons, having been converted from ADU using the gain. Additionally, the averaged frame is normalized by subtracting its mean to aid visualization. The distributions of the normalized averaged frames for both HDR and FFR modes are shown in \autoref{fig:dsnu}. The total, pixel, row and column DSNU was found to be 0.232\,e$^-$, 0.219\,e$^-$, 0.054\,e$^-$ and 0.052\,e$^-$ for the HDR mode and 0.318\,e$^-$, 0.251\,e$^-$, 0.192\,e$^-$ and 0.038\,e$^-$ for the FFR mode. 

Overall the DSNU exhibits minimal variation, though it is more pronounced for the FFR mode. Notably, the FFR mode reveals a bimodal distribution in the row DSNU, as shown in \autoref{fig:dsnu}, which is attributed to the odd/even amplifier pattern also observed in \autoref{fig:row_row}. In FFR mode, row and pixel DSNU are the dominant contributors to the total DSNU, while the column DSNU remains minimal. In contrast, for HDR mode, both row and column DSNU show little variation, with pixel DSNU being the main contributor to the total.

\subsection{Row and Column Noise}
\label{sec:row_column-results}
To investigate the pixel level \textit{temporal} row and column noise in bias frames, we used the same set of (K = 100) bias images at minimum exposure time as described in Section~\ref{sec:bias-results}. These bias frames form a 3-D data cube, where the dimensions are indexed by $k$ (image index), $m$ (row), and $n$ (column). To study the row-wise variation, we compute the average value of all pixels $k$ in each row $m$:

\begin{equation}
\mu_{k,m} = \frac{1}{N} \sum_{n=0}^{N-1} y[k][m][n],
\end{equation}

where N is the number of columns, and $y[k][m][n]$ is the value of the pixel at a particular image. This creates a list of average values for each row m across the frames. The mean value calculated for each list of a row m is then given by:

\begin{equation}
\bar{\mu}_m = \frac{1}{K} \sum_{k=0}^{K-1} \mu_{k,m}.
\end{equation}

Finally, the standard deviation of each row m list across the K images is computed as:

\begin{equation}
\sigma_m = \sqrt{\frac{1}{K-1} \sum_{k=0}^{K-1} \left( \mu_{k,m} - \bar{\mu}_m \right)^2 }.
\end{equation}

In this way, for each of the 2048 rows across 100 frames, we quantify the row temporal noise. The same method is applied to evaluate the column-wise noise. In this case, the averaging is performed across rows rather than columns, and the variation is assessed based on the mean values of each column across all images. This way, the standard deviation of the column temporal noise is calculated.

\begin{figure}
\includegraphics[width=\columnwidth]{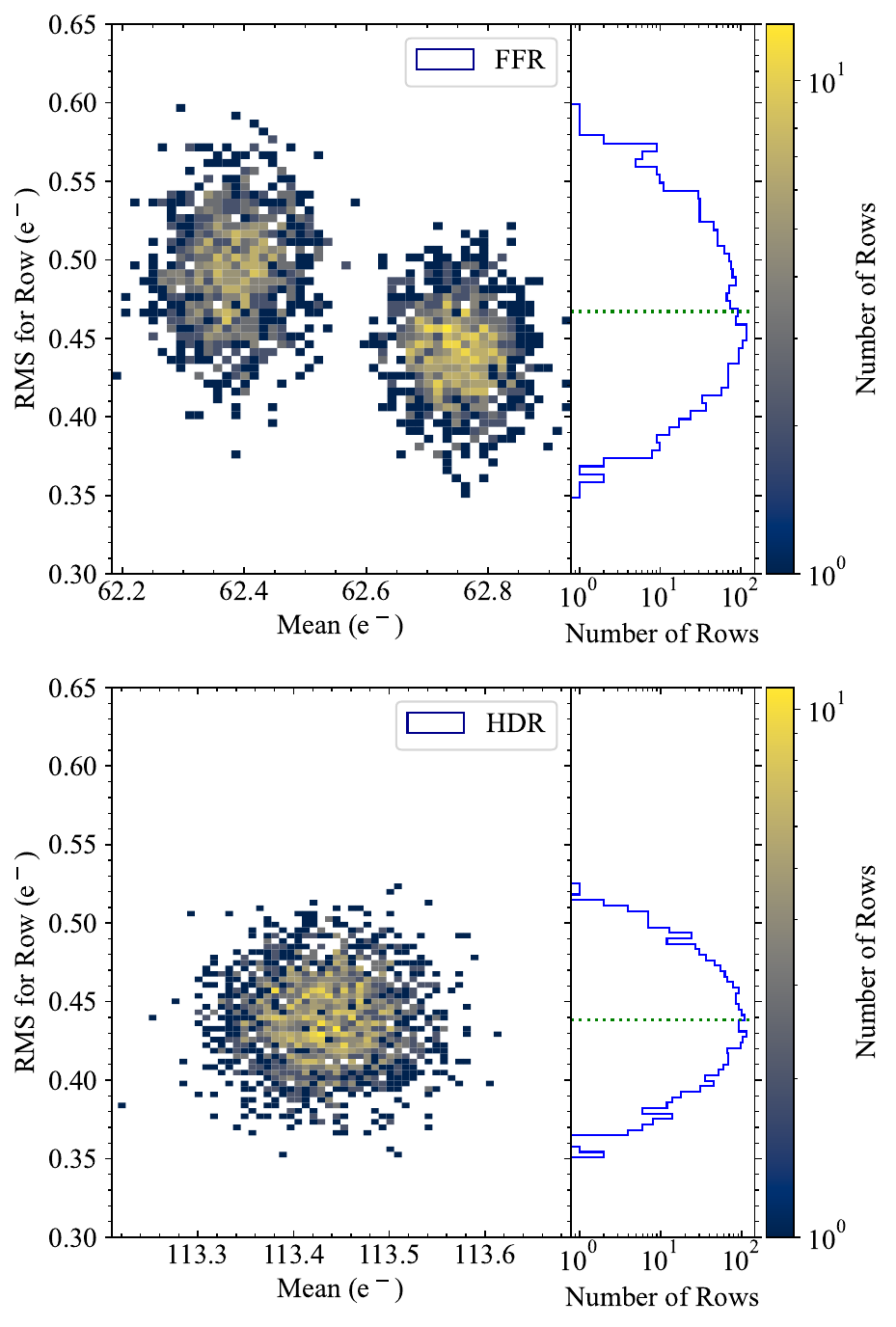}
\caption{Row noise distribution using 100 consecutive bias frames. The 2-D histogram corresponds to row temporal. The scattered plot shows the mean value of each row from these frames while the colour bar indicates the number of rows. The logarithmic row noise distribution is shown in the blue histograms, with the median value as a green dotted line. \textbf{Top:} Mean row noise distribution for the FFR mode. \textbf{Bottom:} Mean row noise distribution for the HDR mode.}
\label{fig:row_row}
\end{figure}

\begin{figure}
\includegraphics[width=\columnwidth]{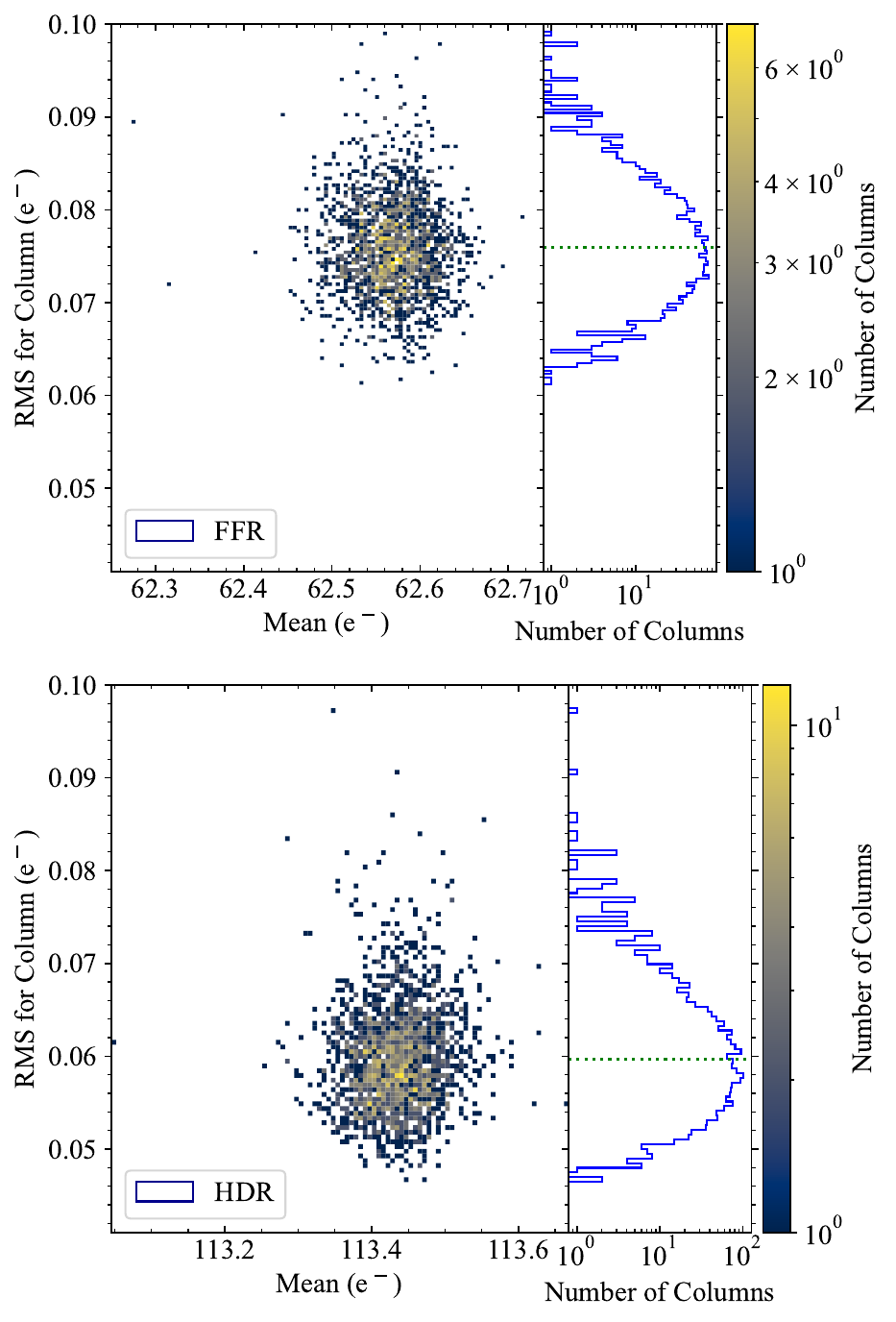}
\caption{Column noise distribution using 100 consecutive bias frames. The 2-D histogram corresponds to column temporal noise. The scattered plot shows the mean value of each column from these frames while the colour bar indicates the number of columns. The logarithmic column noise distribution is shown in the blue histograms, with the median value as a green dotted line. \textbf{Top:} Mean column noise distribution for the FFR mode. \textbf{Bottom:} Mean column noise distribution for the HDR mode.}
\label{fig:column_column}
\end{figure}

Our results plotted with the same format as the read noise plots shown in \autoref{fig:row_row} for the row noise and \autoref{fig:column_column} for the column noise in FFR and HDR mode, respectively. For both modes, the column noise is weaker than the row noise. The mean value for the row noise was measured to be 0.467\,e$^-$ and 0.438\,e$^-$ for the FFR and HDR mode, respectively. The mean value for the column noise was measured to be 0.076\,e$^-$ and 0.06\,e$^-$ for the FFR and HDR mode, respectively.

Both HDR and FFR modes exhibit Gaussian noise distributions; however, in FFR mode, the 2D histogram (top panel of \autoref{fig:row_row}) shows two distinct distribution clusters. This separation is not a result of temporal variation due to row banding but is likely a spatial variation due to the readout architecture in FFR mode, where even rows are processed by the first HG channel and odd rows by the second HG channel (see Section~\ref{sec:modes}). This architecture leads to a noticeable mismatch between the two amplifiers: the first cluster, corresponding to one amplifier, exhibits higher noise of $\sim$0.5\,e$^-$ and a lower bias mean signal of 62.35\,e$^-$, while the second cluster shows lower noise of $\sim$0.43\,e$^-$ and a higher mean value of 62.78\,e$^-$ (99.33\,ADU). The HDR mode is consistent with noise of 0.438\,e$^-$ and a bias mean signal value of 113.45\,e$^-$ (100.3\,ADU) since only the HG channel is implemented for both even and odd rows.

We investigated whether applying a row and column banding correction could improve the overall read noise of the camera. For this, we computed the average value of each row after applying sigma-clipping to exclude the contribution of cosmic rays and hot pixels. The row offsets were then subtracted from the image to remove horizontal banding. The same procedure was repeated for columns to correct vertical banding. An example row and column temporal noise removed frame is show in \autoref{fig:bias_frames_row_col}. Using the row and column noise values, the RMS read noise is expected to be 1.706\,e$^-$ for the HDR and 1.721\,e$^-$ for the FFR mode. However, our measured RMS values with correcting the row-column banding were slightly higher at 1.719\,e$^-$ for HDR and 1.739\,e$^-$ for FFR. This compensation suggests that while banding corrections reduce read noise, additional noise components are present that cannot be removed by this procedure.

\subsection{Dark frame structure and Dark level measurement}
\label{sec:dark_structure}
\begin{figure*}
\includegraphics[width=\textwidth]{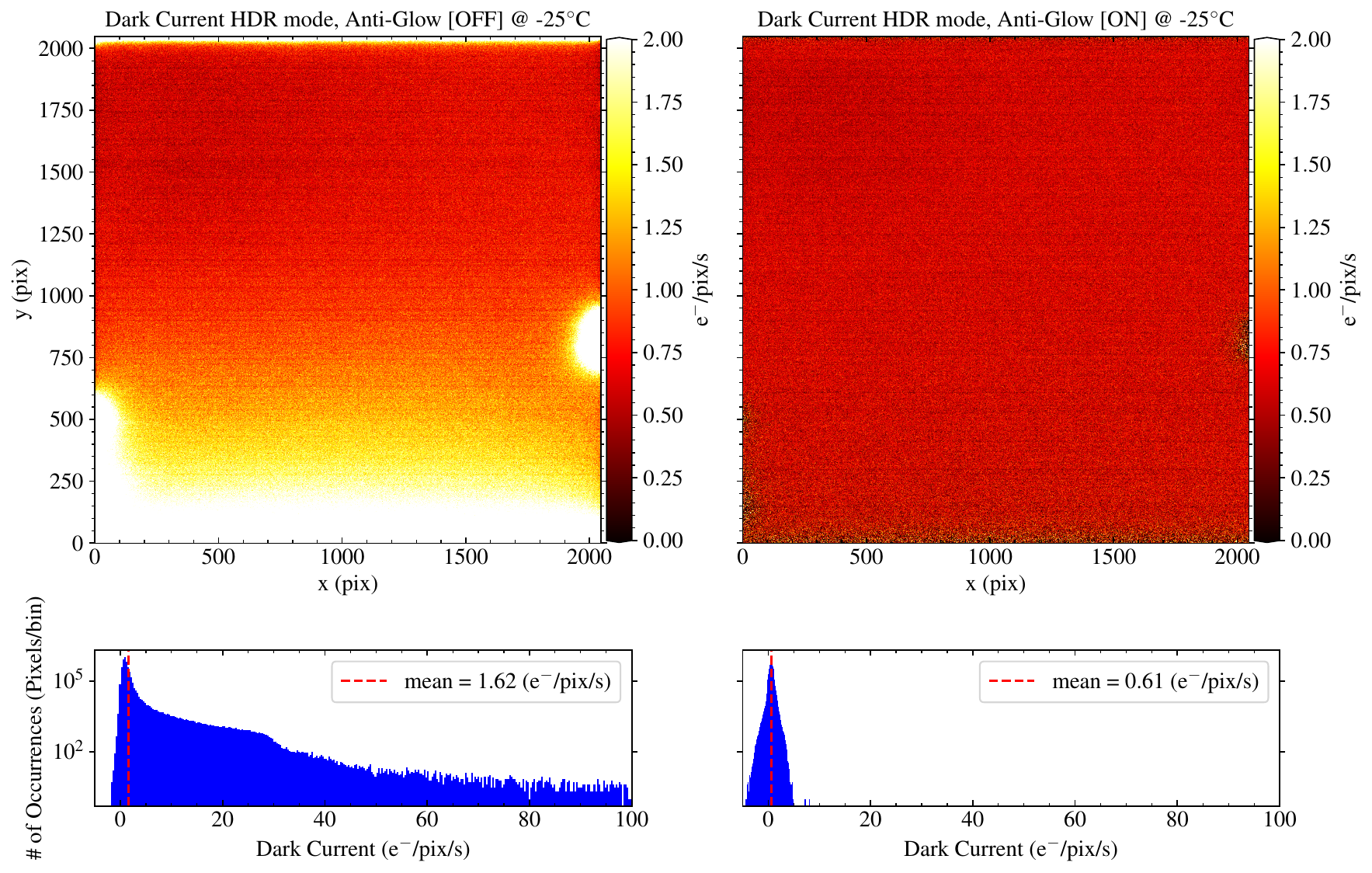}
\caption{Marana dark images taken in the HDR mode at a camera temperature of -25$^\circ$\,C. \textbf{Left} With the Anti-glow correction OFF.  \textbf{Right} With the Anti-glow correction ON. Below each we show the histograms of pixel value distributions from each frame, along with the mean (red dotted line).}
\label{fig:dark_frames_Glow}
\end{figure*}

To study the structure in the dark level, we captured two 10\,s exposure frames at -25$^\circ$C: one with the Anti-Glow correction turned ON and one with the Anti-glow correction turned OFF. Exposures were acquired in HDR mode. To isolate the dark signal, we used the 100 bias frames as in Section~\ref{sec:bias-results}, to construct a master bias image by averaging the pixel values at each position across all frames (see \autoref{eq:DSNU}). This master bias was then subtracted from each dark frame. To verify that the bias subtraction process did not inadvertently introduce noise, we compared the standard deviation of the dark frames with the Anti-Glow correction turned OFF before and after bias subtraction. The measured standard deviation was 3.291,e$^-$ prior to subtraction and 3.298,e$^-$ afterward, indicating no significant change in noise level. Subsequently, each dark frame is divided by the exposure time to convert it to dark level per unit time in \eps. The resulting dark level images, along with histograms of pixel value distribution, are shown in \autoref{fig:dark_frames_Glow}. 

%left image Anti-Glow OFF
The dark image with the Anti-Glow correction OFF shows large scale structure with areas of the image at levels that exceed 2 \eps. Glow spots are evident at the edges of the array and most prominent in the bottom rows, a strip of rows at the top of the array, and spots in the lower left and centre right of the array. These structures are caused by amplifier glow (see Section~\ref{sec:glow}). The full frame image has a mean value of 1.62\,\eps. The presence of glow is also clearly evident in the distribution of the pixels as shown in the histogram of \autoref{fig:dark_frames_Glow}.

%right image Anti-Glow ON
The dark image with the Anti-Glow correction ON is a result of an on-camera master dark frame subtraction for the user requested exposure (see Section~\ref{sec:glow}). This results in the suppression of the large scale glow structure, although the regions of very high glow can still be in the image. Turning the Anti-Glow ON removes the glow pattern from the frame, but will not remove the shot noise produced from the glow. The mean dark level is reduced 0.61 \eps, and the pixel value distribution is much narrower than the distribution with the Anti-glow correction OFF. An improvement in the reference image used for the Anti-glow correction means that our dark image is much more uniform than the dark frame for the Marana presented in \cite{2019AN....340..638K}.  

We note that after the anti-glow correction some pixel values are negative. This phenomenon arises from the short exposure time (10\,s) employed to capture the dark frame. In these cases, the dark current may not be sufficiently large compared to the bias level to be the dominant noise source in the frame. The Poisson noise after the bias level and the pre-computed dark map subtracted remains in the image. As a result, some pixel values will be positive and some negative, centered around a mean of near zero. The negative values are visibly evident in the histogram distributions in \autoref{fig:dark_frames_Glow}. 

The anti-glow correction proves effective in cosmetically removing the glow from the image. Nonetheless, it is essential to note that this correction is a calibration image for addressing glow in dark current images. As such, it does not eliminate the noise component associated with the dark current signal. The glow noise can be reduced if the amplifiers are set in low power mode during exposure and only initiated during the readout time \citep{2007SPIE.6690E..03J, wang20154m}. This is feasible considering the fast speeds of CMOS cameras, where the amplifiers do not have to operate for a long time on a full integration. This approach minimizes glow, leaving dark current as the primary source of measured noise. Nevertheless, this option is not applicable.

In the HDR mode and at -30\,$^\circ$C, the dark current with the Anti-Glow correction ON is 0.50\,\eps, and OFF is 0.80\,\eps, which is in agreement with the value reported by \cite{2019AN....340..638K} at temperature -30\,$^\circ$C. We note that \cite{2024SPIE13103E..0RK} found the dark current for the Sona-11 camera to be 4.5\,\eps at -25\,$^\circ$C. However, it is not clear if this is the average value of the full frame image or in cropped regions, that can potentially be affected by glow.

We see from \autoref{fig:dark_frames_Glow} that the glow is much more prominent in some regions of the image compared with others. Therefore, one way to mitigate the dark current effect is to avoid placing targets of interest within the areas most affected by glow. For photometry, this includes the target star and also the comparison stars.

\begin{figure}
\includegraphics[width=\columnwidth]
{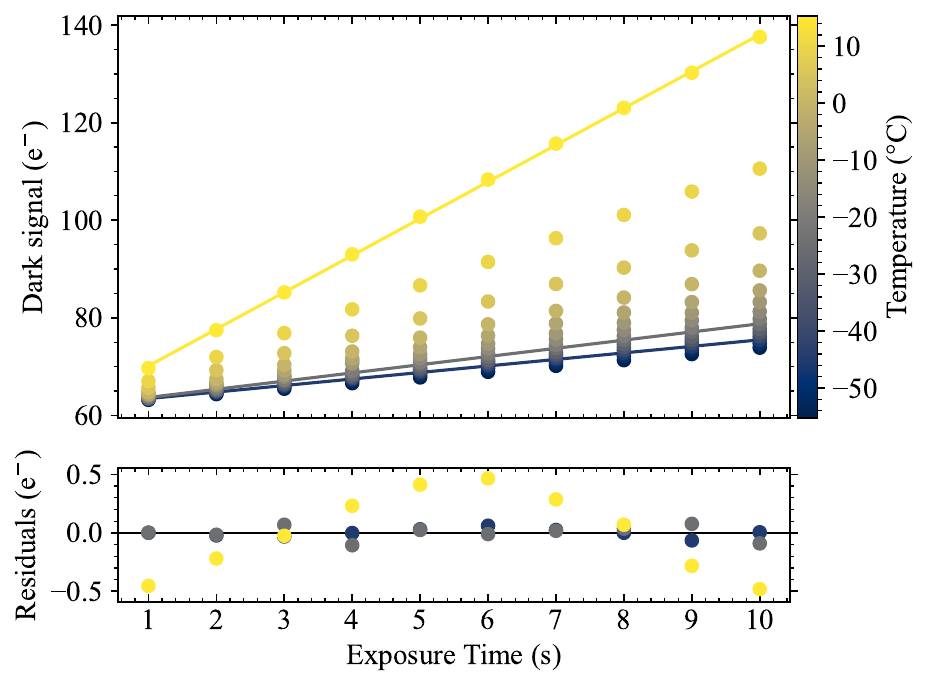}
\caption{Mean dark signal as a function of exposure time over a range of temperatures (color-coded) for the Marana camera in FFR mode. The solid lines represent the slopes at three distinct temperatures at 15\,$^\circ$C, -25\,$^\circ$C and -45\,$^\circ$C. The slope measures the mean dark current of the camera. The intercept indicates the mean bias level of the system. The lower panel shows the residuals from the linear fit from the three temperatures.}
\label{fig:dc}
\end{figure}

%dark signal vs time
To quantify the dark signal for the Marana camera, we measure the mean dark signal as a function of exposure time across a range of temperatures and with the Anti-Glow correction disabled. In order to achieve temperatures below -30\,$^\circ$\,C we used a water cooling system. We take exposures ranging from 1\,s to 10\,s in steps of 1\,s. For each frame we calculate the mean frame value. We used this method for both modes. The results are shown in \autoref{fig:dc} for FFR readout mode. For simplicity, we did not plot all the linear fits and only chose to show the fit for three temperatures at +15\,$^\circ$C, -25\,$^\circ$C and -45\,$^\circ$C\footnote{These are the customer default temperature settings that the camera has been optimized by Andor for best imaging performance.}.

The best fit function to the data shows a very good linear relationship between exposure time and mean pixel level, with typical $R^2$ values for the fit above 0.999 (see \autoref{fig:dc} and \autoref{tab:dark_signal}). The slopes represents the dark current and the y-axis intercept gives the bias level. It is evident that as temperature increases the slope tends to be steeper and therefore increasing the dark current.

For the FFR mode we find that the dark current (dark signal rate) and bias level at -25\,$^\circ$C to be 0.736\,\eps and 98.129\,ADU respectively. For the HDR mode, we find the dark current and the bias level at -25\,$^\circ$C to be 0.672\,\eps and 99.957\,ADU respectively. The bias levels are in agreement with our measurements from Section~\ref{sec:bias-results} yielding an average 98.30$\pm$0.11\,ADU from the intercepts, which is within 1$\sigma$ from the measured bias level for the FFR mode in Section~\ref{sec:bias-results}. The lower panels of \autoref{fig:dc} display the residuals from the linear fit. In both modes, the Marana CMOS camera shows a considerably higher dark current compared to CCDs.

\begin{figure}
\includegraphics[width=\columnwidth]{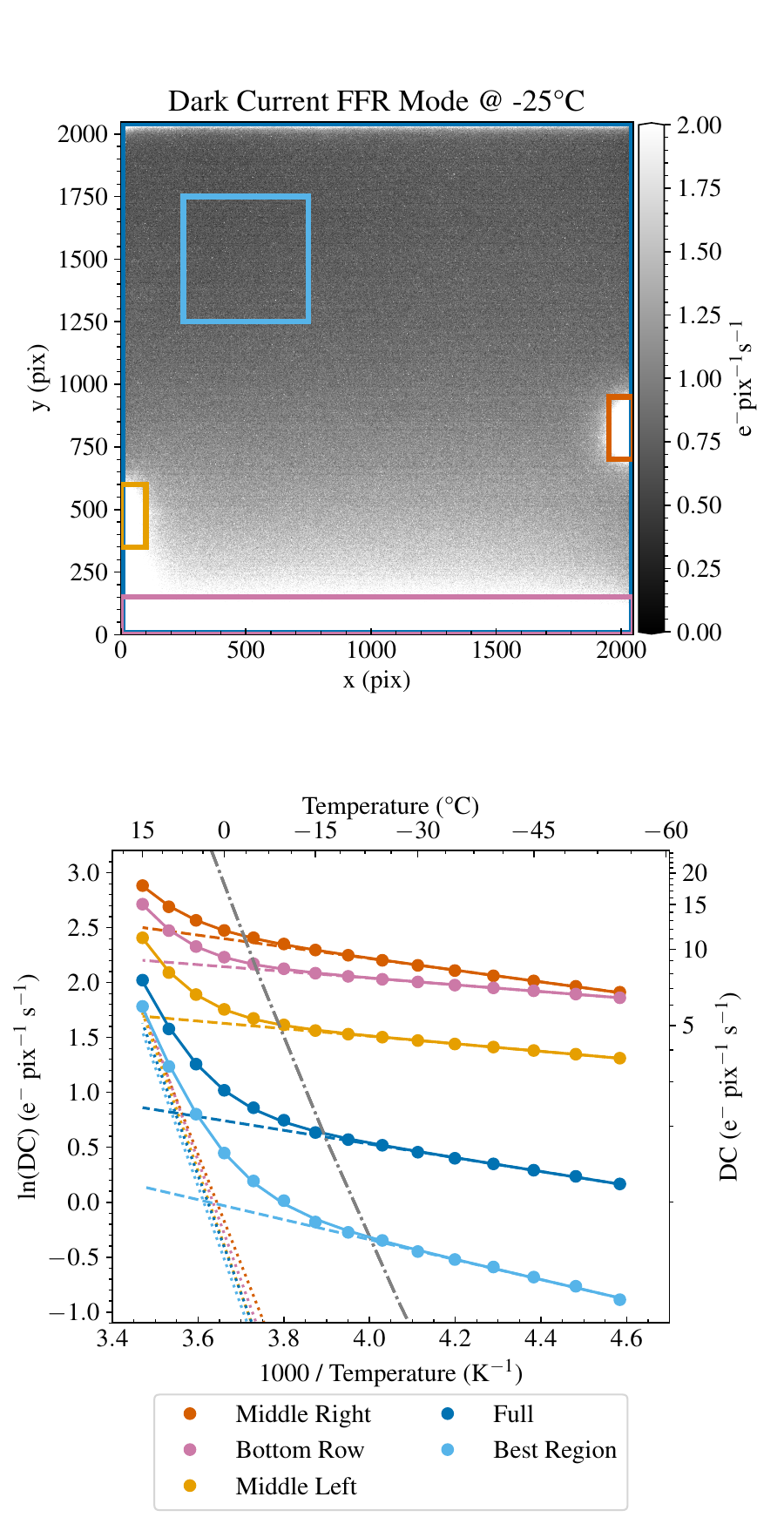}
\caption{The logarithmic dark current as a function of the inverse temperature, following the form of \citet{widenhorn2001meyer}. \textbf{Top:} An example dark current image showing the regions of the used for the measurements. \textbf{Bottom:} The dark current as a function of temperature for each region. The solid coloured lines represent the best-fit model, which is composed of the sum of two exponential functions (shown as the dotted line and the dashed line). The gray dash-dot line and squares represents the dark current of iKon-L SO Blue Sensitive CCD camera from Andor, as reported in the specification sheet.}
\label{fig:dark_frames_T}
\end{figure}

%dark current vs temperature
To investigate the dependence of dark current on camera temperature we used the above analysis for temperature ranging from +15\,$^\circ$\,C to -55\,$^\circ$\,C in steps of 5\,$^\circ$\,C. The slope for each dark signal measurement with the exposures was extracted and plotted in a logarithmic scale \autoref{fig:dark_frames_T}. The predicted dark current ($DC$) follows the Arrhenius behaviour \citep{widenhorn2001meyer} in the form:

\begin{equation}
    \mathrm{DC} = D_0 \exp\frac{-\Delta E}{kT},
\end{equation}

where $D_0$ is a constant, the $\Delta E$ is the activation energy, $k$ is Boltzmann constant and $T$ is the temperature.

As noted by \citet{widenhorn2001meyer}, the dark current can sometimes deviate from a simple exponential trend as a function of temperature. Our data is best fitted using a two-exponential model for the temperature dependence and at different regions on the images. At lower temperatures, the dark current follows an exponential form that is weakly dependent on temperature. At higher temperatures, the data follows an exponential that is more strongly dependent on temperature. To accurately capture this behaviour, we employ a model that is constructed using the following formula:

\begin{equation}
    \mathrm{DC} = D_0 \exp \left(\frac{-A}{T}\right) + D_1 \exp \left(\frac{-B}{T}\right),
\end{equation}

%fix what the values mean
where $D_0$, $D_1$, A and B are constants.

For each region the logarithmic dark current differs, with a tendency of larger values in the glow regions, as shown in the first three plots in \autoref{fig:dark_frames_T}. Overall, our model shows two distinct behaviours with differing dependencies on temperature. At higher temperatures the dark current follows a strong temperature dependency (see values in \autoref{tab:dark_current_slopes}). At lower temperatures the dark current follows a much weaker temperature dependency (see values in \autoref{tab:dark_current_slopes}). The strong temperature dependency, is less evident in the first three plots in \autoref{fig:dark_frames_T}, since those regions are dominated by glow. For the average full frame image measurements, the dark current follows a strong temperature dependency with $A=10.7\,\unit{K}$. At lower temperatures, the dark current follows a much weaker temperature dependency of $B=0.625\,\unit{K}$. The overlapping of the two exponential functions occurs at temperature 8.8\,$^{\circ}$C. The same two-temperature behaviour is observed in all regions, including those with minimal glow. This indicates that the total array is affected by glow, even if the average dark current measured across the full frame is smaller than the values measured directly in the glow dominated regions. For temperatures below 0\,$^\circ$C the dark current is dominated by the amplifier glow, that is weakly dependent on temperature. Only for temperature above 0\,$^\circ$C does the dark current have a significant contribution from the classical thermal noise which is more strongly temperature dependent.

The large dark current signal remains the major drawback of CMOS sensors compared with CCD detectors. For faint objects with low sky background signal, the high dark current from certain CMOS detectors continues to position them as a secondary choice compared to their CCD counterparts. However, some CMOS sensors have significantly less amplifier glow, such as the SONY IMX455, that has demonstrated exceptionally low dark current, even at relatively warm temperatures, reaching as low as 0.018\,\eps at 0\,$^\circ$C \citep{2023sndd.confE...3A}.

\subsection{Photo Response Non-Uniformity (PRNU)}
\label{spatial_prnu}
\begin{figure}
\includegraphics[width=\columnwidth]{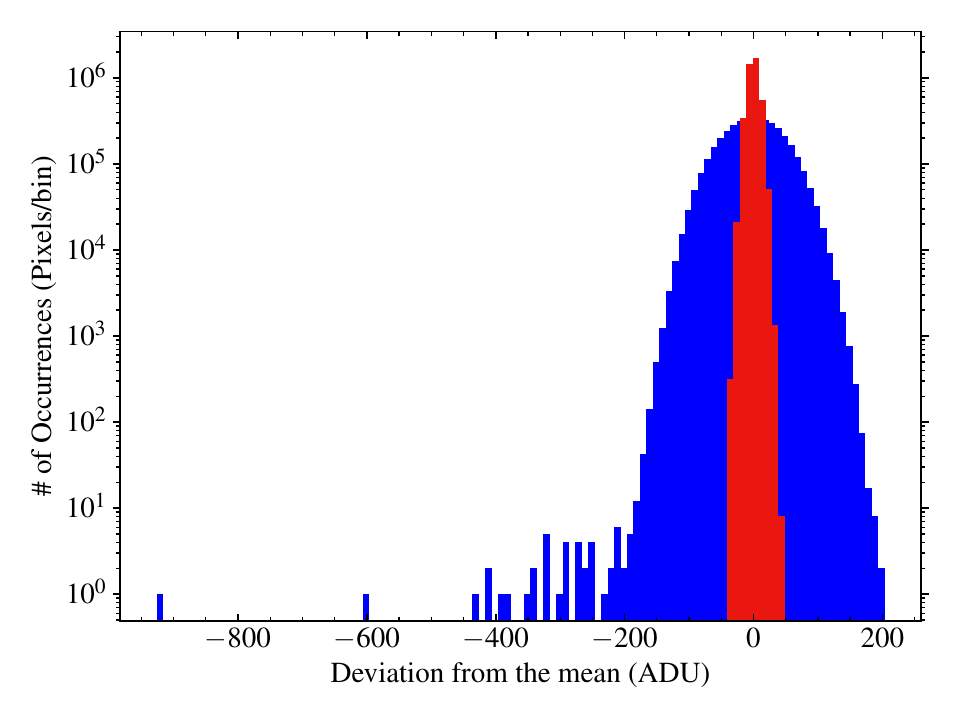}
\caption{Histogram distributions of the photo response non-uniformities for both modes. The red bins indicate the FFR while the blue the HDR mode.}
\label{fig:prnu_dsnu}
\end{figure}

To study the PRNU we follow Section 4 of the EMVA-1288 \citep{jahne2010emva} version 4.0 documentation. The 100 half-saturation images were averaged, and the mean value is calculated as in Section~\ref{sec:bias-results}. To measure the spatial variance we used the following: 

\begin{equation}
s_{y.\mathrm{light}}^2 = \frac{1}{MN}\sum_{m=0}^{M-1}\sum_{n=0}^{N-1}(\Bar{y}_{\mathrm{light}}[m][n]-\mu_{y. \mathrm{light}})^2
\end{equation}

where the spatial variance of the averaged frame is denoted as $s_{y.\mathrm{light}}^2$. However, there is still a temporal variance component included in the value. Therefore, we adjusted the spatial variance by subtracting the residuals of the temporal variance:

\begin{equation}
    s^2_{\mathrm{PRNU}} = s_{y.\mathrm{light}}^2 - \frac{\sigma^2_{m,n.\mathrm{light}}}{K}
\end{equation}

where the variance with the temporal noise suppressed $s^2_{\mathrm{PRNU}}$, the variance before the temporal noise subtraction $s_{y.\mathrm{light}}^2$, the temporal variance $\mathrm\sigma^2_{m,n.\mathrm{light}}$ and K the number of images used. The PRNU is the difference of the variances calculated with and without illumination, and divided by the corrected mean value of the signal at half saturation after subtracting the bias level:

\begin{equation}
    \mathrm{PRNU} = \frac{\sqrt{s^2_{\mathrm{PRNU}} - s^2_{\mathrm{DSNU}}}}{\mu_{y. \mathrm{light}} - \mu_{y.\mathrm{bias}}}
\end{equation}

where the variance of the bias level $s^2_{\mathrm{DSNU}}$, the mean value at half saturation $\mu_{y.\mathrm{light}}$ and the mean bias level $\mathrm{\mu_{y.bias}}$.
The distribution of the PRNU was plotted as shown in the upper panel of \autoref{fig:prnu_dsnu}. The HDR mode exhibits a significantly larger spread compared to the FFR mode. This disparity arises because the half-saturation point for the HDR 16-bit setting is 30000\,ADU, whereas for the FFR 12-bit setting, it is only 2000\,ADU. Temporal suppression was performed using 100 images; however, in HDR mode, the temporal noise still contributes and therefore leads to a broad distribution. In contrast, this effect is not observed in FFR mode, as the lower half-saturation point of 2000\,ADU allows 100 images to sufficiently minimise the spread in the histogram. The results demonstrate that the camera maintains uniform performance, with PRNU values of 0.131\,\% for HDR mode and 0.294\,\% for FFR mode. 

Since the HDR mode operates using the LG channel at half-saturation, the measured PRNU values suggest that the signal is more uniform in the LG channel, as the PRNU in FFR mode is twice that of HDR mode.

\subsection{Linearity Results}
\label{sec:linearity-curve}

The linearity is expressed through the linear relationship between the mean pixel signal and the exposure time with a constant light source and at a fixed temperature. This is accomplished by fitting a linear model using weighted least square regression, following the guidelines outlined in EMVA-1288 version 4.0 \citep{jahne2010emva}. The linear model is described as:

\begin{equation}
    y[i] = mt_{\mathrm{exp}}[i] + b,
\end{equation}

where $y[i]$ is the mean signal from two consecutive frames and  $t_{\mathrm{exp}}$ is the exposure time. Since the bias level has been subtracted the residual offset b is zero. For each mean signal level we assign a weight $1/y[i]^2$. The residuals from the linear regression is measured in percentage and calculated as:

\begin{equation}
    \delta_{y}[i] = \frac{\mu_{y}[i] - (mt_{\mathrm{exp}}[i] + b)}{mt_{\mathrm{exp}}[i] + b} \times 100.
\end{equation}

\begin{figure*}
\includegraphics[width=\textwidth]{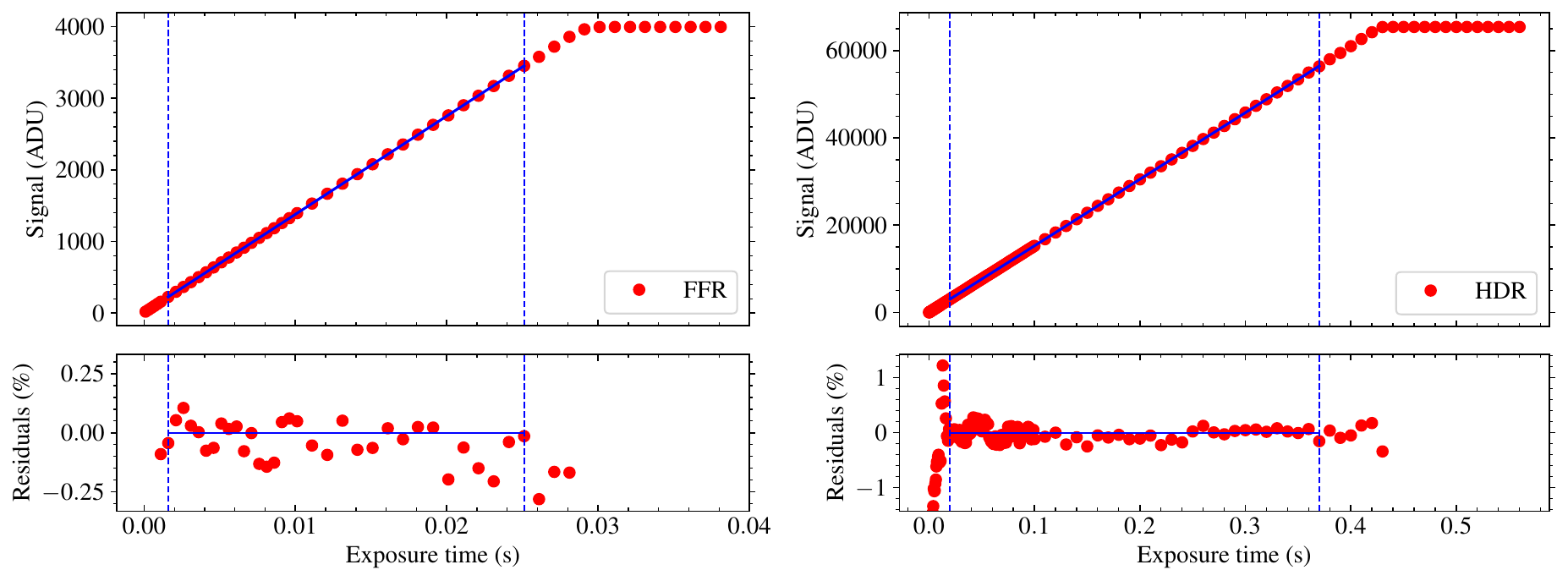}
\caption{Linearity measurements for the Marana in FFR mode (left) and HDR mode (right). The best fit linear model between 5 and 95 \% of saturation is plotted with a blue solid line. Residuals to the linear model are shown in the bottom panels. The dashed horizontal blue lines indicate the 5 to 95\% boundaries we set to measure fit the linearity.}
\label{fig:linearity}
\end{figure*}

We quantify the uncertainty in the linearity using the linearity error ($LE$), which is the average of the absolute values of the deviation of the data points in the range of 5\,\% to 95\,\% from the saturation value:

\begin{equation}
    LE = \frac{1}{n}\sum_{i=1}^{n} \mid \delta_{y}[i] \mid,
\end{equation}

where $n$ is the number of the set of images between 5\,\% and 95\,\% of saturation value. The results for both FFR and HDR modes are shown in \autoref{fig:linearity}. The camera demonstrates small non-linearity, yielding linearity errors of 0.066\,\% and 0.102\,\% for FFR and HDR mode respectively. These results are consistent with Andor specification for linearities within the range of $>99.7$\,\%.

Over the years, the EMVA-1288 standard has introduced updates to the methods used for measuring camera non-linearity. In this study, we also report results obtained using previous methodologies from release version 3.0 and release version 3.1 candidate 1 and candidate 2. The release version 3.0 follows the same approach as release version 3.1 and calculates residuals from linear regression without incorporating weights. Non-linearity is determined as the average distance between the maximum and minimum residual from the linear regression method.

The release version 3.1 candidate 2 uses the weights on the linear regression to estimate the residuals and 
two non-linearity values are reported, one for the maximum and one for the minimum residual values. Therefore, the release version 3.0 yields 0.098\,\% and 0.122\,\% for the FFR and HDR, and release version 3.1 candidate yields maximum 0.105\,\%, minimum -0.205\,\% and maximum 0.277\,\% and minimum -0.250\,\% for the FFR and HDR respectively. 

\begin{figure}
\includegraphics[width=\columnwidth]{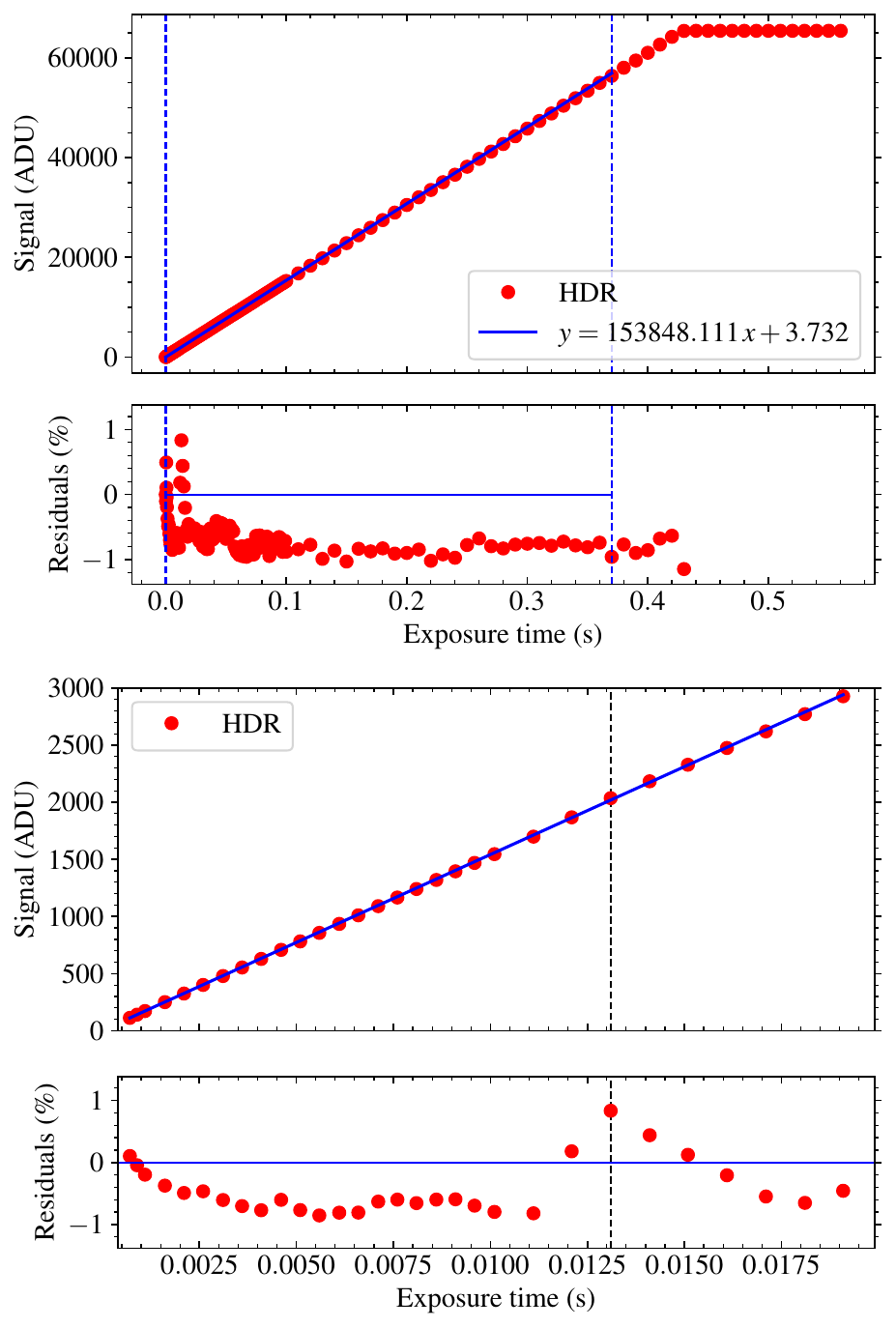}
\caption{\textbf{Top:} Linearity measurements for the Marana in the HDR mode. The best fit linear model between 0.01\,\% and 95\,\% of saturation is plotted with a blue solid line. Residuals to the linear model are shown in the bottom panels. The dashed vertical blue lines indicate the 0.01\,\% to 95\,\% boundaries we set to measure linearity. \textbf{Bottom:} Zoom in panel of the linearity in the transition region. The black dashed line indicates the transition region from HG to LG channel.}
\label{fig:linearity_zoom}
\end{figure}

We note that the linearity is not evaluated below 5\% of saturation when following  the EMVA-1288 standard . Since the transition region in HDR mode occurs below this threshold, it is typically excluded from the calculation. Furthermore, this means that for faint stars with brightness below 5\% of the saturation, linearity is not guaranteed. To investigate this behaviour, we expand the boundaries of the linearity analysis to include the range from 0.01\% to 95\% of saturation. The results are presented in the top panel of \autoref{fig:linearity_zoom}, while the bottom panel provides a close-up view of the transition region between the HG and LG channels. The non-linearity was found to be 0.698\,\%, with maximum and minimum residuals of 0.834\,\% and 1.033\,\% respectively. Those values are larger than the values reported in the range of 5-95\%. 

Despite the transition in the two gain channels when sampling signal levels at $\sim1800$\,ADU, as described in \autoref{fig:Binary_Time_Series}, the camera demonstrates excellent linearity within the range, consistent with the manufacturer's documentation. The Marana camera show good linearity even when measuring it within a wider range of 0.01\,\% to 95\,\%, showing that the sensor is linear in low light signals, despite the $\pm$1\,\% deviation in the transition region, as shown in \autoref{fig:linearity_zoom}. Our linearity measurement is in agreement with previous works in \cite{2021RAA....21..268Q} and \cite{2024SPIE13103E..0RK}.

\subsection{Quantum Efficiency and Transmission Window Results}
\label{sec:QE_window_results}
We measure the photodiode current across a wide range of wavelengths. The responsivity curve of the photodiode, is extracted from the datasheet (see \autoref{fig:PDRE}) and converted to QE values (see \autoref{fig:PDQE} and \autoref{eq:PDQE}). We measure the camera flux by averaging 10 images of the same exposure after subtracting the bias level in the HDR mode. The absolute QE of the camera is defined as:
\begin{equation}
\label{eq:QE_Final}
    QE\,(\%) = \frac{G \times S \times q \times \mathrm{PDQE}}{\mathrm{PDC} \times t} \times 100
\end{equation}

\begin{figure}
\centering
\includegraphics[width=\columnwidth]{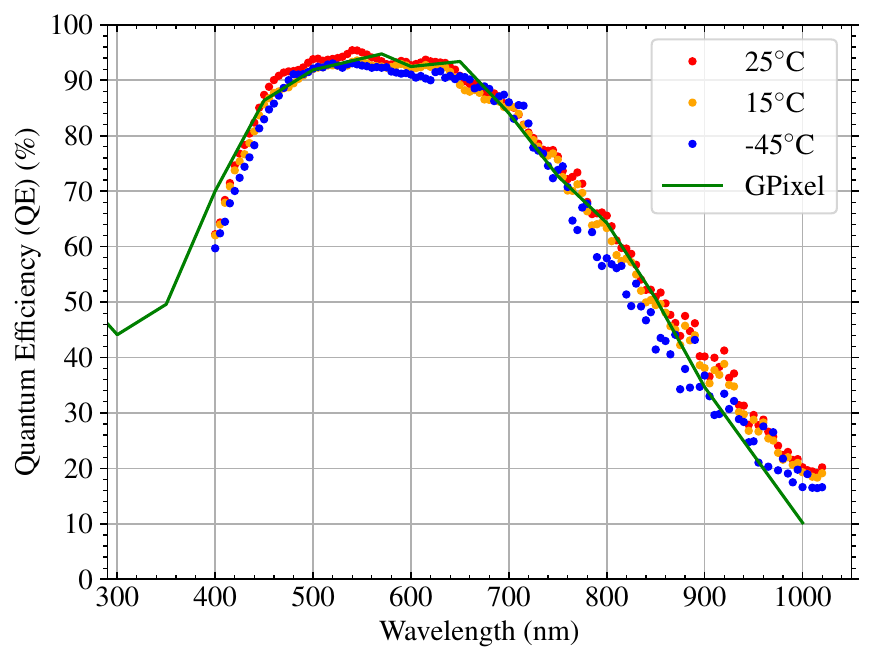}
\caption{QE results for Marana sCMOS camera at three temperatures. With red circles the QE at +25\,$^\circ$C, with orange squares the QE at +15\,$^\circ$C and with blue pentagons the QE at -45\,$^\circ$C. The solid green line represents the QE measured by GPixel.}
\label{fig:MARANA_QE}
\end{figure}

Where $G$ is the gain (e$^-$/ADU), $S$ is the sum of the signal from all pixels in the averaged frame in ADU,  $q$ is the electron charge in Coulomb, PDQE is the QE of the photodiode, PDC is the photodiode current in amperes and $t$ is the exposure time of the images taken with the camera in seconds. The QE was measured at 3 different temperatures; +25\,$^\circ$C, +15\,$^\circ$C and -45\,$^\circ$C as shown in \autoref{fig:MARANA_QE}.

The measured QE is in close agreement with the manufacturer’s specifications within the 450–900\,nm wavelength range. In the 400–450 nm region, the measured QE is up to 10\,\% lower than the manufacturer’s values, whereas in the 950–1020\,nm region, it exceeds the manufacturer’s QE by up to 10\,\%. The Marana camera is most sensitive at $\sim$\,560\,nm with a 95\,\% peak QE. Additionally, the QE curve shows a small increase at higher sensor temperatures and toward the red end of the spectrum when compared against the -45\,$^\circ$C dataset.

Previous studies from \cite{1979JAP....50.1491W} and \cite{velichkointrinsic} show that the QE is red-shifted as the temperature is increasing. This behaviour can likely be explained by the effect of the band gap energy. The band gap energy required to excite an electron on a silicon is 1.12\,$eV$ at constant temperature 300\,$K$ yielding an upper limit of 1107\,nm (cut-off wavelength) \citep{1973JPCRD...2..163S, sze2021physics, stefanov2022cmos} and producing a single electron-hole pair.

\begin{figure}
    \centering
    \includegraphics[width=\columnwidth]{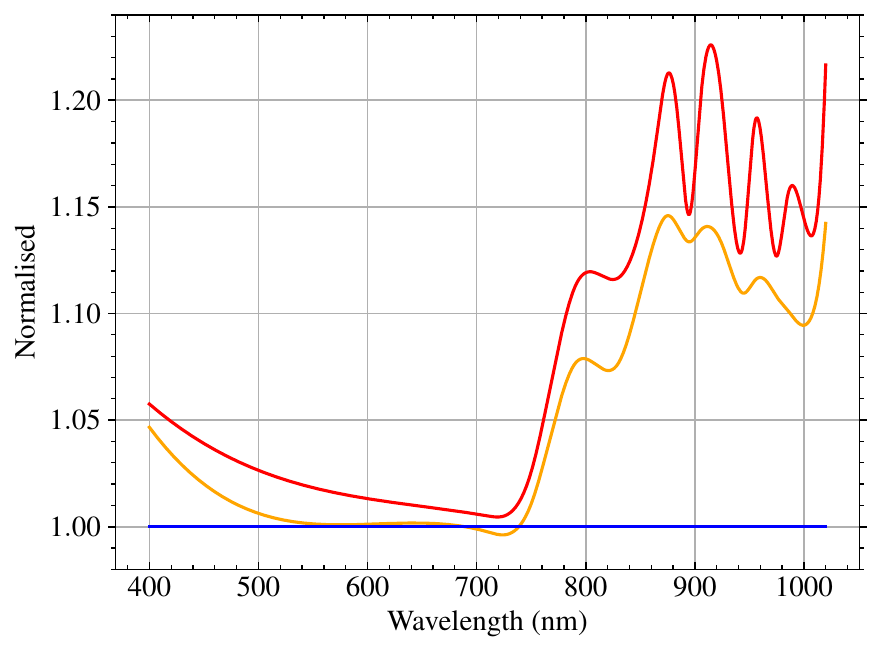}
    \caption{Normalised QE measurements as a function of wavelength with respect to the -45\,$^\circ$C QE measurement, showing with the blue line. Orange line indicates the QE measurement for +15\,$^\circ$C while with red line the QE measurement for +25\,$^\circ$C.}
    \label{fig:QE_T}
\end{figure}

The band gap energy increases slightly at lower temperatures \citep{BENSALEM201755} and therefore the cut-off wavelength is shorter at lower temperatures. Additionally, \cite{2008SEMSC..92.1305G} demonstrated that at lower temperatures the absorption length of the silicon also increases. Consequently, electrons require more energy to transition to the conduction band when incident photons interact with the silicon and therefore fewer electrons are promoted at lower temperatures compared to warmer temperatures. This is expressed as in the following equation from \citep{BENSALEM201755}:

\begin{equation}
\label{eq:QE_T}
    E_g = E_0 - cT
\end{equation}

where the E$_g$ is the band gap energy for silicon, E$_0$ is the band gap energy at 0\,K, T is the temperature in K and c constant. The effect is more evident at longer wavelengths. This is in agreement with the results in \autoref{fig:QE_T} showing the normalised QE measurements with respect to the -45\,$^\circ$C QE measurement, where the normalised ratio increases at redder wavelengths for warmer temperatures. The data is being smoothed using the \texttt{UnivariateSpline} python package.

\begin{figure}
\centering 
\includegraphics[width=\columnwidth]{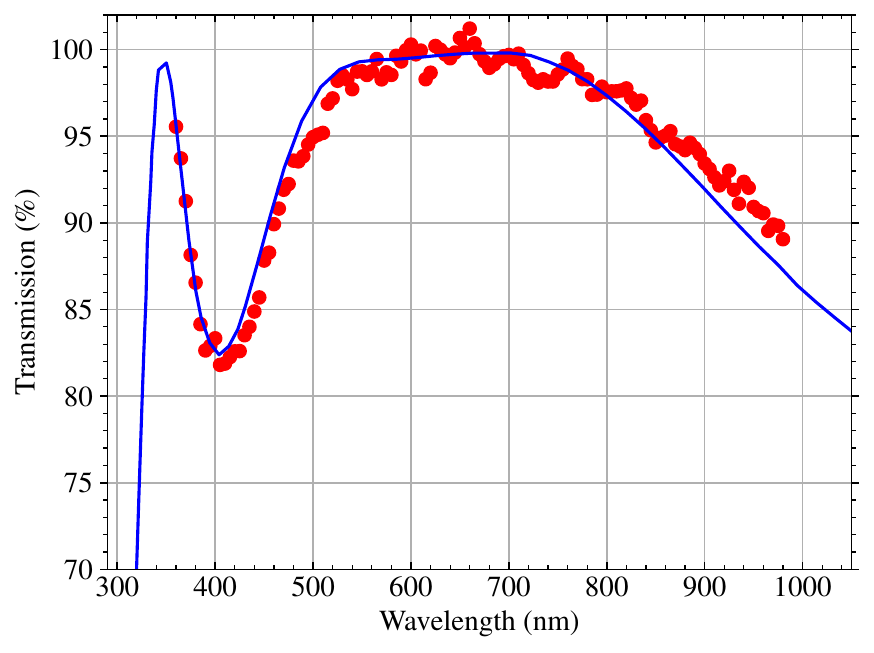}
\caption{Transmission as a function of wavelength for the window which is mounted on the Marana CMOS camera measured as the fractional percentage ratio of the photodiode current measurements from \autoref{fig:PD_Ratio}. The measured data is show with red while for comparison is the transmission curve from Andor with blue solid line. The y-axis is zoomed-in for detailed visualisation.}
\label{fig:Transmission}
\end{figure}

To quantify the transmittance of the window, we measured the amount of photons that successfully pass through the camera’s optical window for various wavelengths. Transmittance is defined as the ratio of the intensities of the transmitted light and the reference light. This measurement shows which photons are efficiently transmitted and which are partially or fully reflected or absorbed by the window material. This is expressed as the following:

\begin{equation}
\label{eq:TRA}
    \mathrm{Tranmission(\lambda)} = \frac{I_{t}(\lambda)}{I_{0}(\lambda)}
\end{equation}

where $I_{t}(\lambda)$ is the photocurrent measured at a specific wavelength with the camera's window in the optical path and $I_{0}(\lambda)$ is the photocurrent measured with a clear path between the light source and the photodiode for the same wavelength. We recorded 10 current measurements for each wavelength step and estimate the mean and the standard deviation. The ratio of the mean values of these current measurements (see \autoref{fig:PD_Ratio}) was plotted as a percentage figure as shown in \autoref{fig:Transmission}. The transmission curve provided by the window supplier has been included in the same figure for comparison. The results show a slight decrease in transmission to approximately 82\,\% in the range of 400-500 nm as illustrated in \autoref{fig:Transmission}. Additionally, Andor reported that the steep increase at wavelengths $<$400\,nm is applied to enhance the sensitivity for applications that require observations at shorter wavelengths. Transmission approaches nearly 100\,\% within the camera’s peak sensitivity range and gradually decreases toward redder wavelengths.

\section{Conclusions}
\label{conclusion}
We present a characterisation of the Andor Marana CMOS camera, according to the EMVA-1288 version 4.0 standard. Laboratory tests were conducted in both HDR and FFR modes of the Marana camera. A summary of the Marana camera performance as measured in this work is set out in \autoref{specs_results}. 

\begin{table*}
\centering
\caption{Specifications of the Marana CMOS camera measured in this work in comparison with a deep depleted back-sided iKon-L 936 CCD which is currently operating at NGTS telescopes.}
\label{specs_results}
\begin{tabular}{lclclclc}
\hline
\hline
\textbf{Camera mode}                           & \textbf{Marana FFR}  &     & \textbf{ Marana HDR} &     & \textbf{ iKon-L}\\ \hline
Digital output                                 & 12-bit        &     & 16-bit       &     & 16-bit\\
Readout modes                                  & 200 MHz       &     & 100 MHz      &     & 3MHz\\
Gain, e$^-$ADU$^{-1}$                          & 0.632         &     & 1.131        &     & 1.99\\
Full Well Capacity, e$^-$                      & 2351          &     & 69026        &     & 82000\\
Non-Linearity, \%                              & 0.099         &     & 0.122        &     & 1\\
Read Noise, Median (RMS) \,e$^-$               & 1.577 (1.785) &     & 1.571 (1.763)&     & 12.9\\
Dark Current Anti-glow ON, \eps           & 0.736@-25$^{\circ}$C         &     & 0.672@-25$^{\circ}$C        &     & N/A\\
Dark Current Anti-glow OFF, \eps          & 1.674@-25$^{\circ}$C         &     & 1.617@-25$^{\circ}$C        &     & 0.003@-80$^{\circ}$C\\
PRNU, \%                                      & 0.294         &     & 0.131        &     & 1.79\\
DSNU, \,e$^-$                                 & 0.318         &     & 0.232        &     & 3.79\\
Row Noise, \,e$^-$                      & 0.467         &     & 0.438        &     & N/A\\ 
Column Noise, \,e$^-$                   & 0.076         &     & 0.06         &     & N/A\\ \hline    
\end{tabular}
\end{table*}

The read noise is low for both modes, and along with the full well depth of the HDR mode, it has good dynamic range. For illumination levels before the transition region, pixels are processed with the HG capacitance yielding low read noise. For illumination levels larger than the transition region, the LG capacitance is employed yielding read noise of 40\,e$^-$. This read noise change may affect stars with shot noise similar to the read noise. The FFR mode exhibits similar read noise characteristics with the HDR mode, however it has low dynamic range. This limitation becomes critical when observing bright stars, which are more likely to saturate in the FFR mode. The FFR mode is more well suited for fast tracking and high cadence photometry of faint stars where saturation is not achieved early.

Since CMOS sensors employ multiple amplifiers, variations between them can result in significant pixel-to-pixel differences in read noise. In addition, individual pixels may be affected by RTS or inherently noisy amplifiers. This behaviour can have a notable impact on the detection of faint stars, where only a few electrons are collected and may become indistinguishable from RTS induced fluctuations. To mitigate these effects, correction maps should be created or existing solutions, such as spurious noise filter, should be applied.

The Marana camera exhibits a dark current level that is higher than what is typically observed in modern CCD sensors (see \autoref{specs_results}, and accompanied by persistently preserved glow spots at the image edges. This contributes significantly to non-uniformities in the frames and the total dark current, creating a blend of dark current and glow. To prevent corruption of the quality of data for high precision photometry, selection of targets near the edges will be avoided as the stars can easily jump out of the frames during tracking. This will result in smaller image format and field of view, which will limited the number of comparison stars for photometry.

Despite the good linearity in the transition region in the HDR mode, faint stars can be affected by the dual amplification from the two channels assuming stars with signal in the same level of the transition. Similar effect for bright stars may occur due to nearby bright pixels in the point-spread function that receive significantly less signal than the star’s central pixels, making them more susceptible to the transition level.

The Marana camera offers several advantages for astronomical applications when compared to CCD-based systems such as the iKon-L (see \autoref{specs_results}). Its fast readout speeds and low read noise can outperform many standard CCD cameras, making it well-suited for high-cadence time-series photometry where improved temporal resolution is required. The absence of a mechanical shutter further reduces operational complexity and minimizes the need for maintenance or complete shutter replacements. However, like most CMOS sensors, including GSENSE400BSI, deep-depletion technology is not applicable yet. As a result, QE at redder wavelengths remains lower compared to deep-depleted CCDs specifically designed for enhanced red sensitivity. Moreover, the Marana camera provides a dynamic range of 93\,dB, which is superior to single-channel cameras such as the QHY600, which utilizes Sony’s IMX series sensors and typically achieves a dynamic range of around 83\,dB \citep{2023sndd.confE...3A}. The GSENSE400BSI sensor used in the Marana also offers a broad QE across the visible spectrum, outperforming the IMX series in this regard and enhancing the camera’s sensitivity.

The next step of our study will be to perform time-series photometry using the CMOS Marana camera in real on-sky conditions. To do this, we plan to mount the Marana on an NGTS \citep{2018MNRAS.475.4476W} telescope at the Paranal observatory in Chile. We plan to run the Marana at NGTS simultaneously with one of the existing iKon-L CCDs, providing a direct on-sky comparison between the CMOS and CCD detectors for time-series photometry.

\section*{Acknowledgements}
This project was conducted as part of a UK Science and Technology Facilities Council (STFC) Industrial CASE (Cooperative Awards in Science and Technology) PhD studentship. IA is the STFC funded PhD student, and gratefully acknowledges the support STFC under the CASE Industry scheme ST/W005077/1 (Project title: Precision Photometry with the new generation of fast readout Scientific CMOS Camera).

This Industrial CASE PhD studentship is collaboration between the University of Warwick and Andor Technology Ltd. The project benefited from a collaborative funding arrangement between the University of Warwick and Andor Technology Ltd as per the Industrial CASE award guidelines. All data analysis, interpretation, and conclusions presented in this work were carried out independently by the Industrial CASE PhD student (IA). Experiments were conducted at the University of Warwick with the exception of the QE experiment, which was conducted at Andor’s laboratories, using optical equipment provided by Andor.

We acknowledge capital funding from the STFC Early Technology Development Capital Funding (ST/W005719/1 Project title:
Discovering New Worlds with sCMOS Cameras). 

We would like to thank the anonymous referees for providing valuable feedback which improved this manuscript.

%%%%%%%%%%%%%%%%%%%%%%%%%%%%%%%%%%%%%%%%%%%%%%%%%%
\section*{Data Availability}
The data the software used is publicly available on the GitHub repository: \url{https://github.com/AperpieGG/Paper_I}. Bias, Dark DSNU and PRNU images used on this work will be made available upon request.
 
%%%%%%%%%%%%%%%%%%%% REFERENCES %%%%%%%%%%%%%%%%%%

% The best way to enter references is to use BibTeX:

\bibliographystyle{rasti}
\bibliography{Revission_III} % if your bibtex file is called example.bib

% Alternatively you could enter them by hand, like this:
% This method is tedious and prone to error if you have lots of references
%\begin{thebibliography}{99}
%\bibitem[\protect\citeauthoryear{Author}{2012}]{Author2012}
%Author A.~N., 2013, Journal of Improbable Astronomy, 1, 1
%\bibitem[\protect\citeauthoryear{Others}{2013}]{Others2013}
%Others S., 2012, Journal of Interesting Stuff, 17, 198
%\end{thebibliography}

%%%%%%%%%%%%%%%%%%%%%%%%%%%%%%%%%%%%%%%%%%%%%%%%%%

%%%%%%%%%%%%%%%%% APPENDICES %%%%%%%%%%%%%%%%%%%%%
\appendix
\section{Bias, Light and temperature stability}
\label{sec:appendix_bias_light_temperature}
\begin{figure}
\includegraphics[width=\columnwidth]{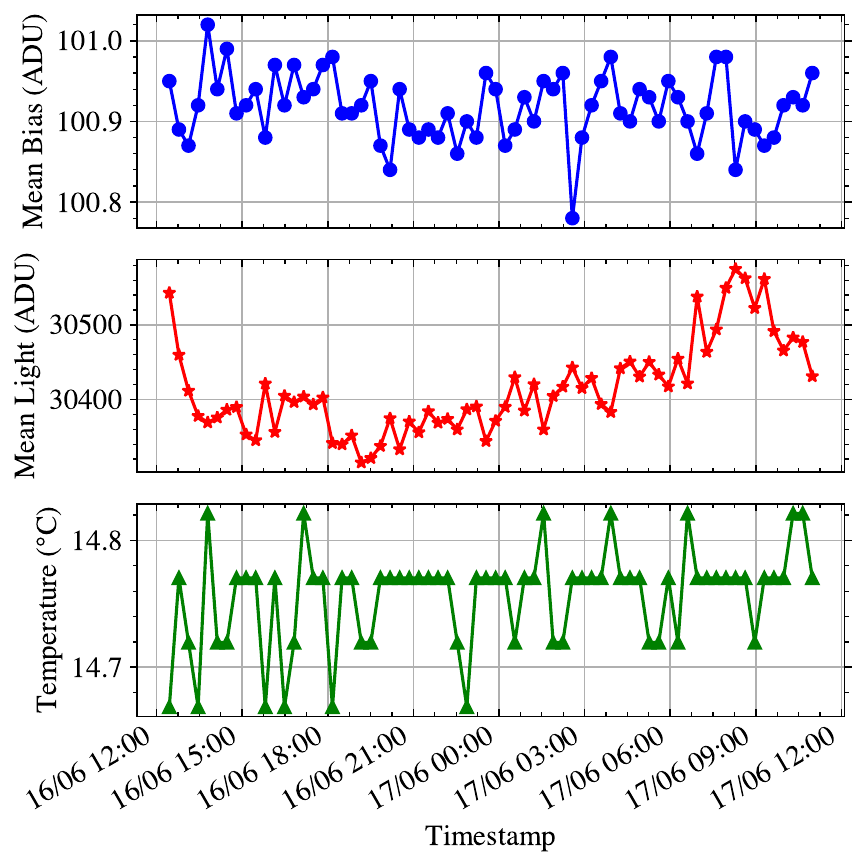}
\caption{Bias level, light source signal and temperature for the Marana camera over time. \textbf{Top:} Average bias level value of a series of images at minimum exposure time and at dark environment. \textbf{Middle:} Average light source signal of a series images at a fixed exposure time. \textbf{Bottom:} Temperature value measured from the camera.}
\label{fig:stability_15}
\end{figure}

\begin{table}
\centering
\caption{Exposures steps based on the saturation exposure (1.5\,seconds.}
\label{tab:exposure_steps}
\begin{tabular}{ccc}
\hline 
\hline
\textbf{Exposure Time}  &       \textbf{Exposure Step}             \\ \hline  
$>=$ 0.0001\,s   &       0.0002\,s     \\
$>=$ 0.001\,s    &       0.0005\,s     \\
$>=$ 0.01\,s     &       0.001\,s      \\
$>=$ 0.1\,s      &       0.01\,s       \\
$>=$ 1\,s        &       0.1\,s        \\ \hline         
\end{tabular}
\end{table}

\begin{figure*}
\includegraphics[width=\textwidth]{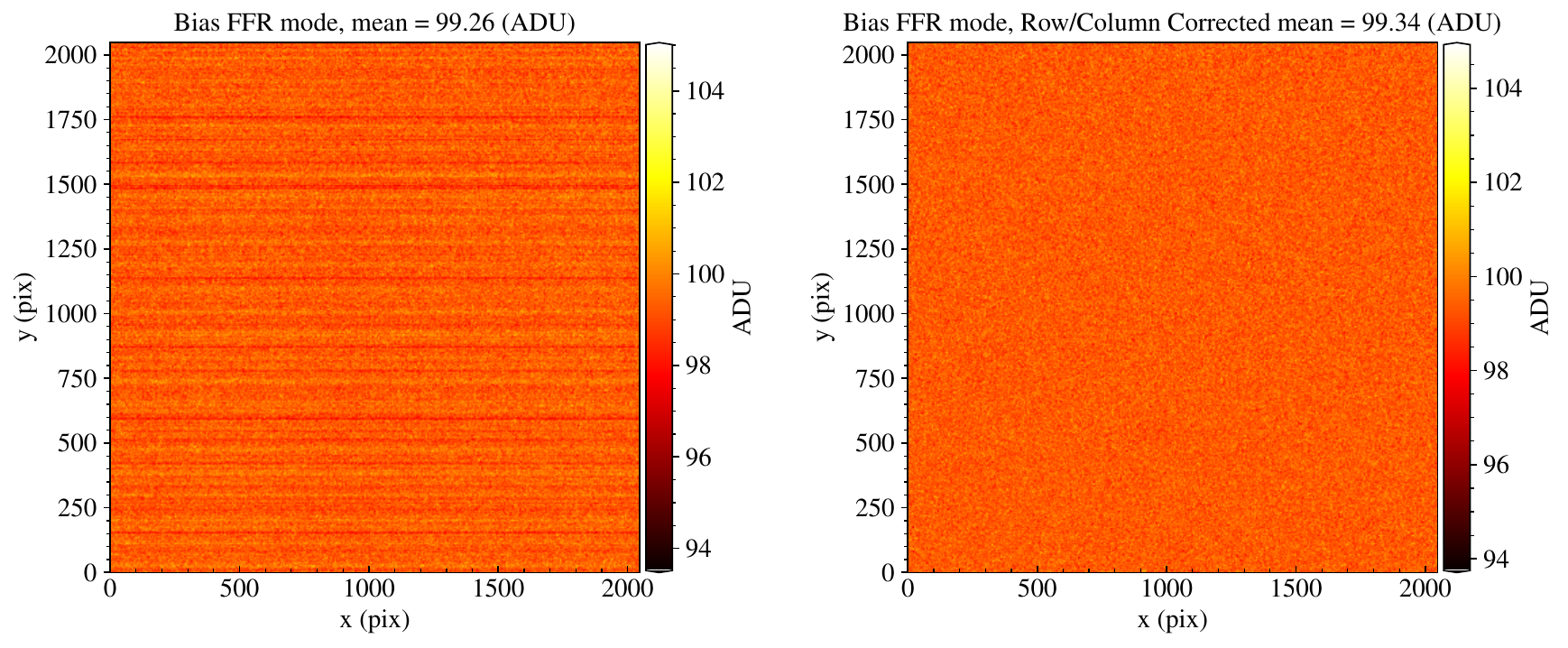}
\caption{Bias frames from Marana camera recorded at minimum exposure for the FFR mode (\textbf{left}) and the FFR mode with the row and column temporal noise removed (\textbf{right}) . The camera is operating at temperature of -25$^\circ$\,C.}
\label{fig:bias_frames_row_col}
\end{figure*}

\newpage
\section{Quantum Efficiency and Transmission Window calculations}
\label{sec:appendix_qe_window}
The QE is expressed by comparing the ratio of the number of electrons with the number of incident photons. This is demonstrated as:

\begin{equation}
\label{eq:QE}
    \mathrm{QE} = \frac{\text{Number of electrons}}{\text{Number of incident photons}}
\end{equation}

\begin{figure}
\includegraphics[width=\columnwidth]
{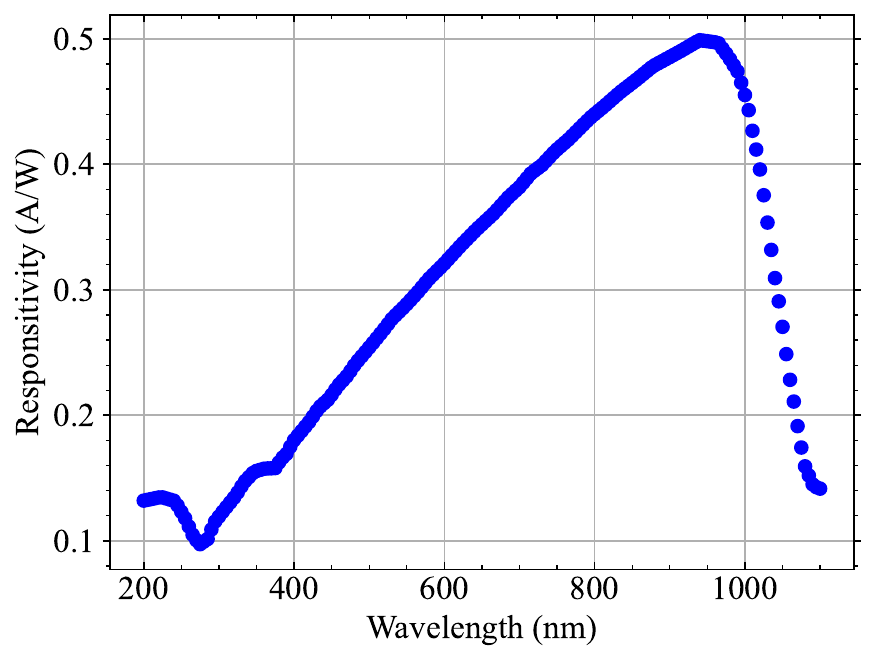}
\caption{The responsivity of the photodiode used for the QE measurements as described in Section~\ref{sec:QE_window_results} plotted as a function of wavelength.}
\label{fig:PDRE}
\end{figure}

HAMAMATSU provides the responsivity of the photodiode, as shown in \autoref{fig:PDRE} which quantifies how efficiently it converts optical power into electrical current. However, since the goal is to compare the QE of the camera with that of the photodiode, the responsivity must first be converted to QE for a direct comparison. From \autoref{eq:QE}, the number of electrons is approximated as: 

\begin{equation}
\label{eq:QE_1}
    N_{e} = \frac{I [A]}{q [C]}
\end{equation}

where the current $I$, measured in Amps divided by the charge of a single electron $q$ in Coulombs. On the other hand, the number of photons striking the photodiode is the light power and is expressed as: 

\begin{equation}
\label{eq:QE_2}
    P [J/s] = N_{p} h [Js] \frac{c [m/s]}{\lambda [nm]}
\end{equation}

Where $P$ is the Power measured in Joules per second, the number of photons $N_{p}$, the Plank's constant $h$ in Joules per second, the speed of light $c$ in meters per second and the wavelength $\lambda$ in nanometres. This equation will take the form: 

\begin{equation}
\label{eq:QE_3}
    N_{p} = \frac{P[J/s] \lambda [nm]}{c [m/s] h [Js]}
\end{equation}

Dividing \autoref{eq:QE_1} with \autoref{eq:QE_3}, yields the QE equation: 

\begin{equation}
\label{eq:QE_4}
    QE = \frac{\frac{I [A]}{q [C]}}{\frac{P[J/s] \lambda [nm]}{c [m/s] h [Js]}}
\end{equation}

And by doing some dimensional analysis to remove the units, leads to: 

\begin{equation}
\label{QE_5}
    QE = \frac{\frac{I}{q}}{\frac{P \lambda \times 10e-9}{ch}}
\end{equation}

Solving mathematically the constants and the remaining is the QE of the photodiode: 

\begin{equation}
\label{eq:PDQE}
    PDQE = 1240 \frac{R}{\lambda}
\end{equation}

\begin{figure}
\includegraphics[width=\columnwidth]
{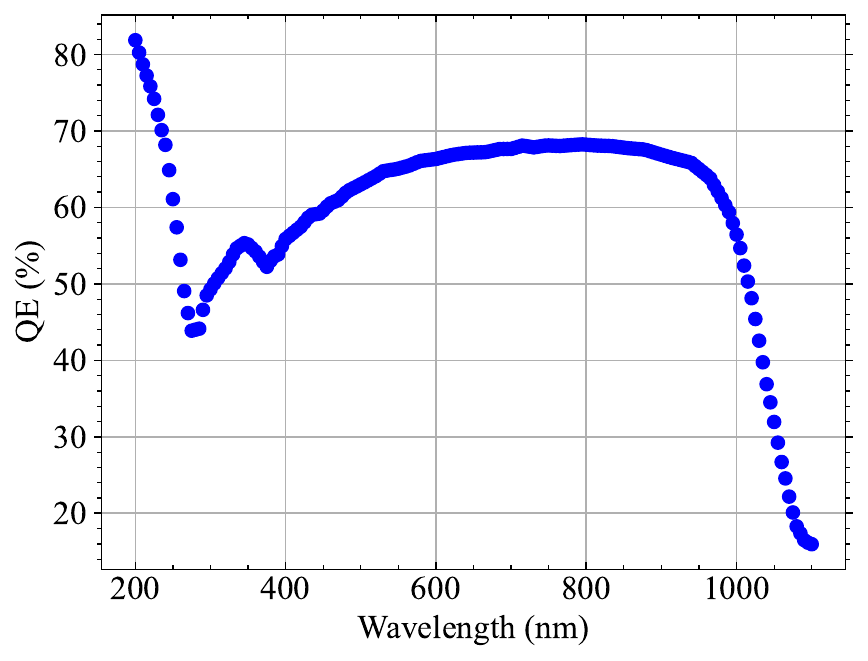}
\caption{The corresponding QE values of the photodiode as a function of wavelength after converting from responsivity using \autoref{fig:PDRE}.}
\label{fig:PDQE}
\end{figure}

Where the responsivity of the photodiode $R$ on a particular wavelength. This equation is employed to convert the responsivity values of the photodiode for each wavelength to QE values as shown in \autoref{fig:PDQE}.

\begin{figure}
\includegraphics[width=\columnwidth]
{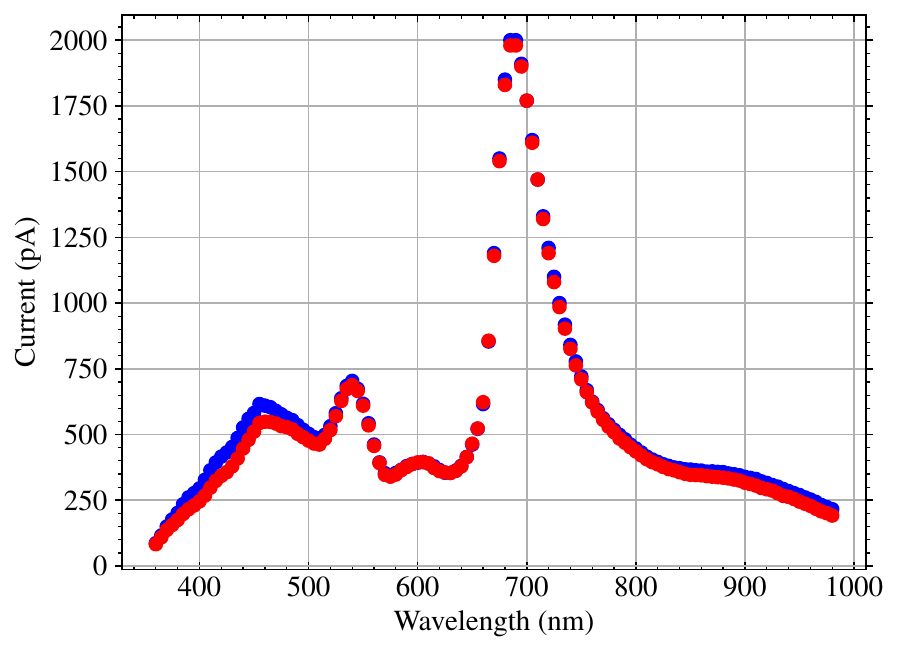}
\caption{Photodiode measurements as a function of wavelength. The current with the transmission window show in red, while with blue current data without the transmission window.}
\label{fig:PD_Ratio}
\end{figure}

\section{Dark signal calculations}
\label{sec:appendix_dark_signal}
\begin{table}
\centering
\caption{Best fit from the linear relation of the dark signal as a function of exposure at different temperatures.}
\label{tab:dark_signal}
\begin{tabular}{cccc}
\hline 
\hline
\textbf{Temperature}  &       \textbf{Best fit}          &  \textbf{$R^2$}   &\textbf{Intercept}    \\ \hline  
-55\,$^{\circ}$C      &       1.18059$\pm$0.00875\,\eps  &   0.99956      &   97.9789\,ADU              \\
-50\,$^{\circ}$C      &       1.26516$\pm$0.00322\,\eps  &   0.99995      &   98.2106\,ADU               \\
-45\,$^{\circ}$C      &       1.33784$\pm$0.00406\,\eps  &   0.99993      &   98.2550\,ADU               \\
-40\,$^{\circ}$C      &       1.41730$\pm$0.00370\,\eps  &   0.99995      &   98.0159\,ADU               \\
-35\,$^{\circ}$C      &       1.49151$\pm$0.00481\,\eps  &   0.99992      &   97.9872\,ADU               \\
-30\,$^{\circ}$C      &       1.57626$\pm$0.00786\,\eps  &   0.99980      &   97.8641\,ADU               \\
-25\,$^{\circ}$C      &       1.67426$\pm$0.00705\,\eps  &   0.99986      &   97.9353\,ADU               \\
-20\,$^{\circ}$C      &       1.76818$\pm$0.00503\,\eps  &   0.99994      &   98.0924\,ADU               \\
-15\,$^{\circ}$C      &       1.88600$\pm$0.04866\,\eps  &   0.99470      &   98.7293\,ADU               \\
-10\,$^{\circ}$C      &       2.10966$\pm$0.00749\,\eps  &   0.99990      &   98.0786\,ADU               \\
-5\,$^{\circ}$C       &       2.36270$\pm$0.01283\,\eps  &   0.99976      &   98.1267\,ADU               \\
0\,$^{\circ}$C        &       2.76896$\pm$0.00598\,\eps  &   0.99996      &   97.9665\,ADU               \\
5\,$^{\circ}$C        &       3.51461$\pm$0.00730\,\eps  &   0.99997      &   98.3077\,ADU               \\
10\,$^{\circ}$C       &       4.84287$\pm$0.01105\,\eps  &   0.99996      &   98.4615\,ADU               \\
15\,$^{\circ}$C       &       7.54922$\pm$0.04071\,\eps  &   0.99977      &   98.8726\,ADU               \\ \hline         
\end{tabular}
\end{table}

\section{Dark current calculations}
\label{sec:appendix_dark_current}
\begin{table*}
\centering
\caption{Best fit of the logarithmic dark current as a function of temperature at different regions on the images.}
\label{tab:dark_current_slopes}
\begin{tabular}{cccc}
\hline 
\hline
\textbf{Region}  &       \textbf{Higher Temperature fit}          &  \textbf{Lower Temperature fit}   &\textbf{Intercept of the fits}    \\ \hline  
Middle Right   & $ln(5.524e+15)$ - 9.948\,$x$   &  $ln(77.30)$ - 0.532\,$x$  &   22.03\,$^{\circ}$C   \\
Bottom Row     & $ln(1.529e+17)$ - 10.886\,$x$  &  $ln(26.20)$ - 0.306\,$x$  &   18.29\,$^{\circ}$C   \\
Middle Left    & $ln(3.123e+17)$ - 11.110\,$x$  &  $ln(17.96)$ - 0.344\,$x$  &   14.76\,$^{\circ}$C   \\
Full           & $ln(6.828e+16)$ - 10.700\,$x$  &  $ln(20.72)$ - 0.625\,$x$  &   8.80\,$^{\circ}$C    \\
Best Region    & $ln(1.035e+17)$ - 10.843\,$x$  &  $ln(27.09)$ - 0.910\,$x$  &   3.70\,$^{\circ}$C    \\ \hline         
\end{tabular}
\end{table*}

% \begin{figure}
% \color{red}
% \includegraphics[width=\columnwidth]
% {Data/DC_Demonstration.pdf}
% \caption{Extended version of \autoref{fig:dark_frames_T}, showing the full-frame dark current as a function of temperature. The blue points represent measurements from the Marana CMOS camera, while the gray dash-dot line corresponds to the iDus 416 CCD camera. The black double-headed arrow indicates the difference in dark current when compared at -90\,$^\circ$C.}
% \label{fig:dc_description}
% \end{figure}

% \section{Some extra material}

% If you want to present additional material which would interrupt the flow of the main paper,
% it can be placed in an Appendix which appears after the list of references.

%%%%%%%%%%%%%%%%%%%%%%%%%%%%%%%%%%%%%%%%%%%%%%%%%%

% Don't change these lines
\bsp	% typesetting comment
\label{lastpage}
\end{document}